%% file: main.tex
\documentclass[11pt]{article}
\PassOptionsToPackage{table}{xcolor}
\PassOptionsToPackage{hypertexnames=false}{hyperref}
\usepackage{amsmath}
\usepackage{amssymb}
\usepackage{amsthm}
\usepackage{cite}
\usepackage{astaple-polyu-template}
\usepackage{xurl}
\usepackage{caption}

\usepackage{algorithm}
\usepackage{algpseudocode}
\usepackage{array}
\usepackage{booktabs}
\usepackage{colortbl}
\usepackage{makecell}
\usepackage{multirow}
\usepackage{placeins}
\usepackage{pifont}
\usepackage{subcaption}
\usepackage{tabularx}
\usepackage[most]{tcolorbox}
\usepackage{tikz}
\pgfdeclarelayer{surveybackground}
\pgfsetlayers{surveybackground,main}
\usepackage[edges]{forest}
\usepackage{xspace}
\usepackage{wasysym}

\newcommand{\fullcirc}{\CIRCLE}
\newcommand{\halfcirc}{\LEFTcircle}
\newcommand{\emptycirc}{\Circle}
\newcolumntype{L}[1]{>{\raggedright\arraybackslash}p{#1}}
\newcolumntype{C}[1]{>{\centering\arraybackslash}p{#1}}
\newcolumntype{Y}{>{\raggedright\arraybackslash}X}

\definecolor{stageblue}{HTML}{315FA4}
\definecolor{stagecyan}{HTML}{137F8C}
\definecolor{stagepurple}{HTML}{6F5A9A}
\definecolor{softblue}{HTML}{F1F5FA}
\definecolor{stagegreen}{HTML}{277A68}
\definecolor{stagered}{HTML}{B34C46}
\definecolor{stageamber}{HTML}{A4752E}
\definecolor{tablehead}{HTML}{E8EFF7}
\makeatletter
\newcommand{\tablecolorleft}[1]{\cellcolor{#1}%
  \g@addto@macro\CT@cell@color{\@tempdimb=0pt\relax}}
\newcommand{\tablecolorright}[1]{\cellcolor{#1}%
  \g@addto@macro\CT@cell@color{\@tempdimc=0pt\relax}}
\makeatother
\definecolor{treeink}{HTML}{24364B}
\definecolor{treeline}{HTML}{7D8FA6}
\definecolor{treeleaf}{HTML}{F4F0F8}
\definecolor{treeblue}{HTML}{EBF1F8}
\definecolor{treegreen}{HTML}{EEF6F2}
\definecolor{treegold}{HTML}{FAF4E8}
\definecolor{posdelta}{RGB}{0,120,0}
\definecolor{negdelta}{RGB}{180,0,0}

\setASTAPLETitle{Machine Unlearning for Large Language Models: Foundations, Advances, and Agentic Extensions}
\setASTAPLEAuthors{%
Xiaoyu Xu\textsuperscript{1}, Minxin Du\textsuperscript{1*},
Li Bai\textsuperscript{2}, Junxu Liu\textsuperscript{1},
Yaxin Xiao\textsuperscript{1}, Kun Fang\textsuperscript{1},
Liu Yang\textsuperscript{1} \\ Huadi Zheng\textsuperscript{3},
Peizhao Hu\textsuperscript{3}, Qingqing Ye\textsuperscript{1},
Haibo Hu\textsuperscript{1*}}
\setASTAPLEAffiliation{%
\textsuperscript{1}The Hong Kong Polytechnic University, Hong Kong, China\\
\textsuperscript{2}Southeast University, Nanjing, China
\textsuperscript{3}Huawei Technologies, Shenzhen, China\\
\textsuperscript{*}Corresponding authors}
\setASTAPLEEmail{%
\href{mailto:xiaoyu0910.xu@connect.polyu.hk}{xiaoyu0910.xu@connect.polyu.hk}\\
\href{mailto:minxin.du@polyu.edu.hk}{minxin.du@polyu.edu.hk}\quad
\href{mailto:haibo.hu@polyu.edu.hk}{haibo.hu@polyu.edu.hk}}
\setASTAPLEAbstract{
Machine unlearning aims to remove target influence while preserving other capabilities. This survey compares methods, benchmarks, and evidence across large language models and systems using retrieval, memory, tools, and interacting agents. A five-layer framework connects removal requests, system boundaries, target locations, interventions, and supported claims. A seven-stage lifecycle and six evidence dimensions guide comparison. The review shows that target construction, retained data, and recovery tests affect reported outcomes. Evidence from model evaluations remains insufficient to establish removal across external state and subsequent updates, motivating evaluation that tracks dependencies and tests whether target influence returns.
}

\begin{document}
\ASTAPLEMakeTitle
\ASTAPLEMakeAbstract

\input{sections/01_introduction}

\input{sections/03_foundations_boundaries}
\input{sections/04_lifecycle_taxonomy}
\input{sections/05_standalone_llms}
\input{sections/06_augmented_llms}

\input{sections/07_agents_multiagent}
\input{sections/08_evaluation_verification}
\input{sections/09_applications_deployment}
\input{sections/10_trends_evidence}
\input{sections/11_open_challenges}
\input{sections/12_conclusion}

\bibliographystyle{plainurl}
\bibliography{references}
\end{document}

%% file: sections/01_introduction.tex
\section{Introduction}
\label{sec:introduction}


Large language models (LLMs) have evolved from standalone text generators into the reasoning engines of AI systems. Modern LLM applications augment parametric knowledge with external retrieval, persistent memory, and tool use, enabling access to up-to-date information, long-term interaction context, and external services~\cite{nips/LewisPPPKGKLYR020,acl/LuoTCLLKKYM26,icml/ChengA25}. Building on these capabilities, LLM-based agents can plan, act, observe feedback, and coordinate multi-step tasks, while multi-agent systems further distribute information and decision-making across interacting agents and shared environments.

\input{tables/survey_comparison}
However, this evolution also amplifies privacy, copyright, and security risks. At the LLM level, web-scale training may lead models to memorize and reproduce personal, copyrighted, or outdated information~\cite{iclr/CarliniIJLTZ23,emnlp/KaramolegkouLZS23,corr/Cheng24}. These concerns have already prompted regulatory and legal action: Italy's data-protection authority imposed a EUR~15 million fine on OpenAI in 2024, although the decision was later set aside following a successful court challenge, while several newspapers have sued OpenAI and Microsoft over the alleged unauthorized use of copyrighted content~\cite{garante/OpenAI24,news/APNewspapersOpenAI24}. The risks expand as LLMs become personalized agents that retain conversations and memories while accessing files, calendars, search histories, and external services~\cite{openai/MemoryFAQ26,google/GeminiPersonalization25,anthropic/ClaudeMemory26}. EchoLeak further demonstrated how access to connected applications can turn a crafted input into cross-application data exfiltration in a production agentic system~\cite{aaais/ReddyG25}.

Reliable information removal is therefore a security and governance requirement for both LLMs and agents. In LLMs, sensitive or copyrighted information may persist in model parameters and remain recoverable through generation or extraction attacks~\cite{emnlp/KaramolegkouLZS23,uss/CarliniTWJHLRBS21,acl/WangHZX0TH25}. In agentic systems, the same information may additionally survive in retrieval corpora, vector indexes, long-term memories, caches, tool states, and interaction trajectories, including clinical records and personal histories~\cite{corr/Liao26,iclr/WuWYZCY25}. Such distributed persistence complicates compliance with erasure and deletion rights under privacy frameworks such as the GDPR's ``Right to be Forgotten,'' the CCPA, and PIPEDA~\cite{nips/GinartGVZ19,harding2019,opc_canada_online_reputation_2018}. \emph{Machine unlearning} provides a principled approach that aims to remove the target information while preserving utility on retained data~\cite{nips/GinartGVZ19,sp/BourtouleCCJTZL21}. Although originally formulated for trained models, its scope must now extend from removing information from LLM parameters to controlling its persistence and subsequent use throughout agentic systems.

Existing unlearning studies differ in their targets, intervention scope, and evaluation protocols. Targets range from training records and entity-specific facts to learned capabilities, such as using a designated tool~\cite{emnlp/hong24,icml/ChengA25}. Interventions may modify model parameters, regulate retrieval-conditioned behavior, or jointly update parameters and memory~\cite{corr/Wang2410RAG,icml/MuresanuTZP25,corr/Wang2602Agentic}. However, suppressing a response, deleting a record, and preventing that record from influencing subsequent actions are distinct outcomes, each requiring appropriate evidence. A comparison must therefore connect each study's removal target to the state it modifies and the outcome its evaluation tests within the scope.

\subsection{Scope and Positioning}
Prior surveys examine both the removal of information from trained models and the role of forgetting in learning and adaptation. The former review removal methods and their evaluation~\cite{journals/csur/XuZZZY24,journals/tist/NguyenHRNLYN25}, while the latter consider how forgetting can benefit subsequent learning~\cite{journals/csur/ShaNH26}. Within the removal literature, LLM surveys classify interventions on parameters, inputs, and outputs, distinguish removal from suppression, and examine training, post-training, and inference~\cite{air/BlancoJusticiaJMSDCT25,natmi/LiuYJCBHYLXLVBKL25,corr/Geng2503,corr/ren2506,air/LeKhacT25,corr/Qiu2510}. Table~\ref{tab:survey-comparison} identifies a less developed aspect of this coverage: how model-level unlearning is coordinated with removal from retrieval stores, caches, agent memory, and shared state. These components can retain or reintroduce target information after a model update. Consequently, evidence of removal from a model alone does not establish that a deployed system satisfies the same request.

\begin{figure*}[!t]
  \centering
  \input{figures/agentic_evolution}
    \caption{Overview of unlearning across standalone LLMs, augmented LLMs, LLM agents, and multi-agent systems. An authorized removal request applies across these scopes, with representative information loci, interventions, and removal requirements shown for each. The stepped layout illustrates expanding state and dependencies. The five-layer framework connects request semantics, system scope, information locus, intervention, and evidence-backed claims to provide a common basis for comparing studies. The seven-stage lifecycle organizes execution and maintenance, while the six evidence dimensions guide evaluation within the declared claim boundary. Maturity indicators qualitatively summarize available evidence.}
  \label{fig:evolution}
\end{figure*}
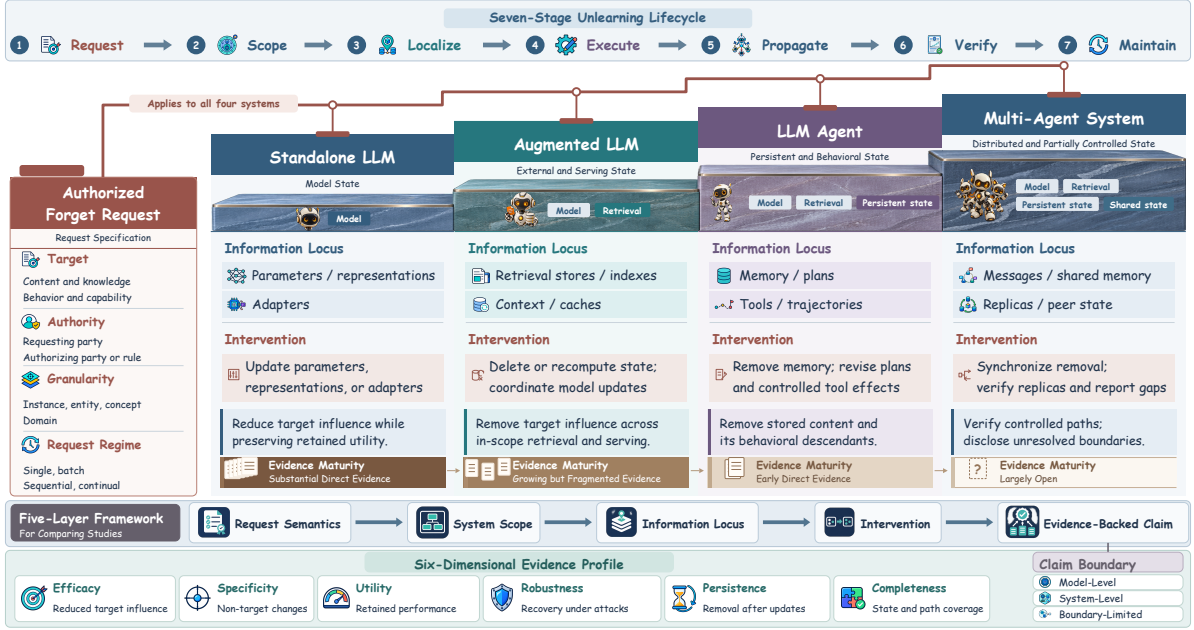

To address this gap, we broaden our review beyond standalone LLMs to examine unlearning in augmented LLMs, LLM agents, and multi-agent systems, where retrieval, persistent memory, tool use, and information sharing introduce additional pathways through which target information can persist or affect behavior. Specifically, we examine which components can be modified and verified, how removal propagates between them, and whether its effects persist after subsequent updates. These settings require different interventions: deleting a record, changing an answer, and revoking tool access affect different state and therefore require different tests. For each intervention, we examine what the reported tests establish about target removal within the stated scope. We also draw on research on supporting mechanisms and security or systems constraints to identify dependencies and limitations that those tests may not cover. This distinction concerns how evidence supports a claim, rather than assigning studies to mutually exclusive categories. We also extend the discussion to forgetting used to improve learning or adaptation, termed \emph{functional forgetting}, whose benefits must be distinguished from evidence that the requested target's influence has been removed.

To organize the comparison of removal methods across these system scopes, we propose a five-layer framework that connects removal requirements to the conclusions supported by evaluation. The layers follow a sequence of questions. \emph{\textcircled{1} Request semantics} asks what information is authorized for removal and what must be preserved, establishing the requirements against which a method is assessed. \emph{\textcircled{2} System scope} asks which components fall within those requirements and which can be modified or inspected. Within this scope, \emph{\textcircled{3} Information locus} asks where the target is stored or can influence behavior, identifying the state an intervention must address. \emph{\textcircled{4} Intervention} asks what the method changes at these locations and how those changes address the request. Finally, \emph{\textcircled{5} Evidence-backed claim} asks what the reported tests establish about the requested removal, given their coverage and assumptions. By proceeding from requirements to system components, target locations, interventions, and evidence, the framework makes explicit what each method addresses and which conclusions remain unsupported. This provides a common basis for explaining why similar interventions can support different claims under different requests and system boundaries.

Figure~\ref{fig:evolution} connects the framework to a seven-stage lifecycle and six evidence dimensions. The framework organizes the comparison of studies, while the lifecycle describes how removal is carried out through \emph{Request}, \emph{Scope}, \emph{Localize}, \emph{Execute}, \emph{Propagate}, \emph{Verify}, and \emph{Maintain}. The evidence dimensions organize the tests used to assess the resulting claim: \emph{efficacy} concerns reduction of the target's influence; \emph{specificity} concerns preservation of unrelated information; \emph{utility} concerns preservation of retained capabilities; \emph{robustness} concerns resistance to recovery attempts; \emph{persistence} concerns continued removal after updates and interactions; and \emph{completeness} concerns coverage of the required components and paths. Together, these views connect the requested outcome, the removal process, and the available evidence without treating results from one model checkpoint as proof of removal throughout a deployed system.

 
\textbf{Our main contributions are summarized as follows:}

\begin{enumerate}

\item \textbf{A structured review of methods and evidence.}
We synthesize research through 25 August 2026, distinguishing direct removal evidence from supporting mechanisms and adjacent research. We compare methods and benchmarks across a range of settings, examining how target construction, retained utility, privacy, recovery tests, and score aggregation shape the conclusions drawn from these studies.

\item \textbf{A unified five-layer framework.}
We connect request semantics, system scope, information locus, intervention, and evidence-backed claims to provide a common basis for comparing unlearning across standalone LLMs, augmented LLMs, LLM agents, and multi-agent systems within a single analytical framework.

\item \textbf{An unlearning lifecycle and evidence profile.}
We connect seven stages, from Request to Maintain, with six evidence dimensions: efficacy, specificity, utility, robustness, persistence, and completeness. This distinguishes the steps needed to fulfill a request from the evidence needed to support a removal claim, while keeping access, cost, and guarantee strength separate throughout the full request handling process.

\item \textbf{Open challenges grounded in evidence gaps.}
We identify open problems in coordinating removal across model parameters, external stores, and agent memory, and in verifying that target influence does not return after later updates under the specified removal requirements. These directions follow from the gap between what existing studies demonstrate and what removal across a deployed system requires.

\end{enumerate}

\input{sections/02_survey_methodology}

%% file: tables/survey_comparison.tex
\begin{table*}[!h]
\caption{Coverage of existing surveys across core topics, system scopes, and assurance, including request processing, maintenance, and supported-claim boundaries. Eval.: evaluation; MAS: multi-agent systems.}
\label{tab:survey-comparison}
\centering
\fontsize{7}{8.5}\selectfont
\setlength{\tabcolsep}{1.5pt}
\renewcommand{\arraystretch}{1.18}
\begin{tabularx}{\textwidth}{@{}L{0.19\textwidth}C{0.05\textwidth}*{8}{>{\centering\arraybackslash}X}@{}}
\toprule
\multirow{2}{*}{\textbf{Review}} & \multirow{2}{*}{\textbf{Year}}
& \multicolumn{3}{c}{\textbf{Core topics}}
& \multicolumn{3}{c}{\textbf{LLM system scopes}}
& \multicolumn{2}{c}{\textbf{Assurance}} \\
\cmidrule(lr){3-5}\cmidrule(lr){6-8}\cmidrule(lr){9-10}
& & \makecell{\textbf{Request}\\\textbf{semantics}}
& \makecell{\textbf{Method}\\\textbf{taxonomy}}
& \makecell{\textbf{Eval.}\\\textbf{criteria}}
& \makecell{\textbf{Retrieval}\\\textbf{context}}
& \makecell{\textbf{Agent}\\\textbf{state}}
& \makecell{\textbf{MAS}\\\textbf{state}}
& \makecell{\textbf{Unlearning}\\\textbf{lifecycle}}
& \makecell{\textbf{Claim}\\\textbf{boundary}} \\
\midrule
\multicolumn{10}{@{}c}{\textit{General machine unlearning}} \\
\addlinespace[2pt]
Xu et al.~\cite{journals/csur/XuZZZY24} & 2024
& \fullcirc & \fullcirc & \fullcirc
& -- & -- & --
& \halfcirc & \fullcirc \\
Nguyen et al.~\cite{journals/tist/NguyenHRNLYN25} & 2025
& \fullcirc & \fullcirc & \fullcirc
& -- & -- & --
& \fullcirc & \fullcirc \\
Sha et al.~\cite{journals/csur/ShaNH26} & 2026
& \fullcirc & \fullcirc & \fullcirc
& -- & -- & --
& \halfcirc & \fullcirc \\
\midrule
\multicolumn{10}{@{}c}{\textit{LLM-oriented unlearning reviews}} \\
\addlinespace[2pt]
Blanco-Justicia et al.~\cite{air/BlancoJusticiaJMSDCT25} & 2025
& \fullcirc & \fullcirc & \fullcirc
& \halfcirc & \emptycirc & \emptycirc
& \halfcirc & \halfcirc \\
Liu et al.~\cite{natmi/LiuYJCBHYLXLVBKL25} & 2025
& \fullcirc & \fullcirc & \fullcirc
& \halfcirc & \emptycirc & \emptycirc
& \halfcirc & \fullcirc \\
Geng et al.~\cite{corr/Geng2503}$^{\dagger}$ & 2025
& \halfcirc & \fullcirc & \fullcirc
& \halfcirc & \emptycirc & \emptycirc
& \emptycirc & \halfcirc \\
Ren et al.~\cite{corr/ren2506}$^{\dagger}$ & 2025
& \fullcirc & \fullcirc & \fullcirc
& \halfcirc & \emptycirc & \emptycirc
& \halfcirc & \fullcirc \\
Le-Khac and Truong~\cite{air/LeKhacT25} & 2025
& \halfcirc & \fullcirc & \fullcirc
& \halfcirc & \emptycirc & \emptycirc
& \halfcirc & \halfcirc \\
Qiu et al.~\cite{corr/Qiu2510}$^{\dagger}$ & 2025
& \fullcirc & \fullcirc & \fullcirc
& \halfcirc & \emptycirc & \emptycirc
& \halfcirc & \halfcirc \\
\midrule
\rowcolor{softblue}
\textbf{This survey} & 2026
& \fullcirc & \fullcirc & \fullcirc
& \fullcirc & \fullcirc & \fullcirc
& \fullcirc & \fullcirc \\
\bottomrule
\end{tabularx}
\par\smallskip
\begin{minipage}{\textwidth}
\fontsize{6.5}{8}\selectfont
\fullcirc\ Dedicated section, taxonomy, or comparison;\quad \halfcirc\ Limited discussion;\quad \emptycirc\ No substantive treatment identified;\quad -- Not assessed;\quad $^{\dagger}$ Preprint. 
\end{minipage}
\end{table*}

%% file: figures/agentic_evolution.tex
\includegraphics[width=\textwidth]{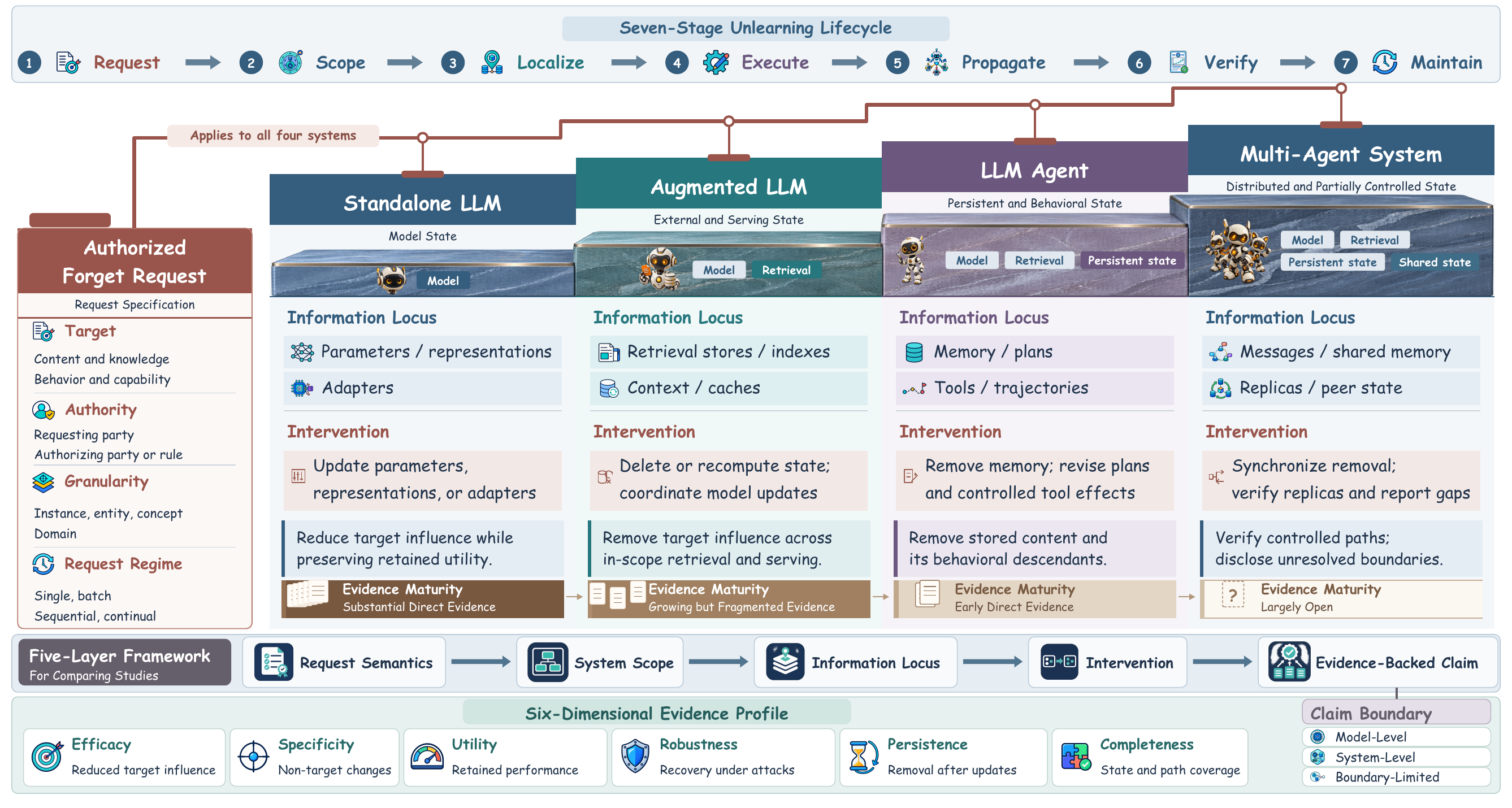}

%% file: sections/02_survey_methodology.tex
\textbf{Organization.} Sections~\ref{sec:foundations}-\ref{sec:framework} introduce the foundations and develop the unified framework. Sections~\ref{sec:standalone}-\ref{sec:augmented} review methods and evidence across standalone LLMs, augmented LLMs, and agent systems. Sections~\ref{sec:system-evaluation}-\ref{sec:applications} discuss how unlearning is verified and applied, including the distinct role of functional forgetting. Section~\ref{sec:trends} describes the review methodology and compares findings across studies. Section~\ref{sec:challenges} examines open challenges and future directions; Section~\ref{sec:conclusion} concludes.

%% file: sections/03_foundations_boundaries.tex
\section{Foundations and Problem Formulation of LLM Unlearning}
\label{sec:foundations}





Deleting a training record, suppressing an answer, and clearing an agent's memory can produce different outcomes. Comparing such interventions requires a shared account of what should be removed, which state is affected, and what evidence demonstrates success. This chapter establishes that account before we apply it to existing methods. Section~\ref{sec:classical-llm-unlearning} introduces classical and LLM unlearning; Section~\ref{sec:state-requests-control} defines requests and system boundaries; Section~\ref{sec:claim-definitions} distinguishes intervention outcomes and their evaluation; and Section~\ref{sec:unlearning-attack-surface} explains the authorization and security conditions that constrain removal. Table~\ref{tab:notation} collects the notation for requests, affected state, interventions, and evaluation.

\begin{table*}[!ht]
\caption{Shared notation for requests, system state, interventions, and evaluation.}
\label{tab:notation}
\centering
\small
\setlength{\tabcolsep}{4pt}
\renewcommand{\arraystretch}{1.08}
\arrayrulecolor{stageblue!65}
\begin{tabularx}{\textwidth}{@{}L{0.31\textwidth}Y@{}}
\toprule
\rowcolor{stageblue!16}
\tablecolorleft{stageblue!16}\textbf{Symbol} & \tablecolorright{stageblue!16}\textbf{Definition} \\
\midrule
\tablecolorleft{stageblue!8}$\mathcal{D},\mathcal{D}_f,\mathcal{D}_r$ & Training data, forget subset, retained complement \\
\rowcolor{stageblue!4}
\tablecolorleft{stageblue!8}$\mathcal{A},\mathcal{U},\mathcal{M}_{\theta}$ & \tablecolorright{stageblue!4}Training algorithm, unlearning operator, model with parameters $\theta$ \\
\tablecolorleft{stageblue!8}$\theta,\theta_u,\theta_{-f}$ & Original, unlearned, retrained-reference parameters \\
\rowcolor{stageblue!4}
\tablecolorleft{stageblue!8}$\mathcal{S}_t=(\mathcal{C}_t,\mathcal{E}_t,\sigma_t)$ & \tablecolorright{stageblue!4}System state: components, dependency edges, state assignment \\
\tablecolorleft{stageblue!8}$r=(T_r,a_r,g_r,\rho_r,B_r,\tau_r,H_r)$ & Target, authority, granularity, regime, boundary, policy, horizon \\
\rowcolor{stageblue!4}
\tablecolorleft{stageblue!8}$B_r^{\mathrm{ctrl}}\subseteq B_r$ & \tablecolorright{stageblue!4}Directly controllable and inspectable requested state \\
\tablecolorleft{stageblue!8}$\mathrm{Cl}_t(r),\mathcal{P}_r$ & Required state closure, declared influence paths \\
\rowcolor{stageblue!4}
\tablecolorleft{stageblue!8}$t_u,\mathcal{S}_{t_u+h}^{u}$ & \tablecolorright{stageblue!4}Intervention completion time, evolved system state at offset $h$ \\
\tablecolorleft{stageblue!8}$\mathcal{A}_r,\mathcal{H}_r$ & Evaluated access/attack conditions, time offsets from $t_u$ \\
\rowcolor{stageblue!4}
\tablecolorleft{stageblue!8}$V_{p,a,h},\epsilon_p$ & \tablecolorright{stageblue!4}Residual target score, maximum admissible value \\
\tablecolorleft{stageblue!8}$Q_h,\eta_r$ & Retained-utility score, minimum admissible value \\
\rowcolor{stageblue!4}
\tablecolorleft{stageblue!8}$D_h,\kappa_r$ & \tablecolorright{stageblue!4}Non-target deviation score, maximum admissible value \\
\tablecolorleft{stageblue!8}$z_p,b_p,w_p$ & Test-completion flag, residual-pass flag, positive path weight \\
\rowcolor{stageblue!4}
\tablecolorleft{stageblue!8}$\mathcal{E}_r$ & \tablecolorright{stageblue!4}Six-dimensional evidence profile; Section~\ref{sec:claim-definitions} \\
\bottomrule
\end{tabularx}
\end{table*}

\subsection{Classical and LLM Unlearning}
\label{sec:classical-llm-unlearning}

Let $\mathcal{D}$ denote the training dataset and $\mathcal{A}$ a randomized training algorithm that produces parameters $\theta=\mathcal{A}(\mathcal{D})$. We write $\mathcal{M}_{\theta}$ for the model parameterized by $\theta$. For a training-data removal request, $\mathcal{D}_f\subseteq\mathcal{D}$ is the forget subset, and $\mathcal{D}_r=\mathcal{D}\setminus\mathcal{D}_f$ is the retained subset. An unlearning operator produces updated parameters $\theta_u=\mathcal{U}(\theta,\mathcal{D}_f;\mathcal{I}_r)$, where $\mathcal{I}_r$ denotes available auxiliary information, such as retained data or training artifacts. Retraining on the retained subset produces reference parameters $\theta_{-f}=\mathcal{A}(\mathcal{D}_r)$ and the corresponding reference model $\mathcal{M}_{\theta_{-f}}$. Exact unlearning requires equality of the distributions of $\theta_u$ and $\theta_{-f}$ over the algorithms' randomness, not agreement on a finite prompt set. Approximate guarantees relax this distributional criterion under specified assumptions~\cite{nips/GinartGVZ19,sp/BourtouleCCJTZL21,nips/SekhariAKS21,nips/GuptaJNRSW21,alt/Neel0S21}.

LLM requests need not identify a training subset: facts or capabilities may derive from unknown records, and post-deployment memory may never have entered training. Such requests require an explicit target and evaluation reference. A convenient forget dataset can omit contributing records, while refusal alone does not establish removal. Membership inference and extraction reveal residual information missed by average accuracy~\cite{sp/ShokriSSS17,sp/CarliniCN0TT22,uss/CarliniTWJHLRBS21,uss/MeeusJRM24}, and memorization varies across examples and models~\cite{iclr/CarliniIJLTZ23,iclr/ShiAXHLB0Z24}. Differential privacy bounds training-data influence under its assumptions; it does not execute a later authorized removal request against designated information after deployment~\cite{jpc/DworkMNS16}.

Most optimization-based methods seek a low value of a forget-retain surrogate with an optional proximity penalty,
\begin{equation}
J_r(\vartheta)=\lambda_f L_f(\vartheta;T_r)
             +\lambda_r L_r(\vartheta;\mathcal{D}_r)
             +\gamma\Omega(\vartheta,\theta),
\qquad L_r(\vartheta;\mathcal{D}_r)=\mathbb{E}_{x\sim\mathcal{D}_r}[\ell(x;\vartheta)],
\label{eq:unlearning-objective}
\end{equation}
where $L_f$ is a method-specific removal loss, $\ell$ a retained-task loss, $\Omega$ a proximity penalty, and the weights $\lambda_f$, $\lambda_r$, and $\gamma$ are nonnegative. Expectations use declared empirical data or sampling distributions. When retained data are unavailable, the method must identify a proxy for $\mathcal{D}_r$ or omit $L_r$. Gradient-ascent unlearning uses $L_f=-\mathbb{E}_{x\sim\mathcal{D}_f}\ell(x;\vartheta)$; preference and representation methods define different removal losses~\cite{acl/YaoCDNWCY24,colm/ZhangLBM24,icml/LiPGYBGLDGMHLJL24,icml/GuoSSED25}. A finite run may stop before convergence. Knowledge editing revises selected associations through targeted updates~\cite{aaai/LiWSKQCWL26}. Module-based interventions can subtract selected parameter components associated with unwanted behavior~\cite{aaai/HuLHZLZ24}. State deletion removes targeted stored entries, such as raw memories and summaries~\cite{corr/Lei2606Agents}. These interventions extend beyond $J_r$.

\subsection{State, Requests, and Control Boundaries}
\label{sec:state-requests-control}

We represent an application at time $t$ by $\mathcal{S}_t=(\mathcal{C}_t,\mathcal{E}_t,\sigma_t)$, where $\mathcal{C}_t$ contains its components and $\sigma_t$ assigns their states. Components may include models, adapters, context, corpora, indexes, caches, memories, tools, and agents. Directed edges in $\mathcal{E}_t$ describe copying, derivation, synchronization, or possible behavioral influence between components; they do not establish causation. This representation distinguishes where information resides from how it may spread or affect behavior, allowing interventions on either an entire component or a target-bearing record within it.

A requester submits a removal request, an authorizing party approves it, and an operator executes it; one party may hold several roles. The request $r=(T_r,a_r,g_r,\rho_r,B_r,\tau_r,H_r)$ specifies target, authority, granularity, regime, scope, policy, and horizon. Policy $\tau_r$ governs retention, future use, and authorized reintroduction. Scope $B_r$ identifies required components and state, including external copies. The operator may modify and inspect only $B_r^{\mathrm{ctrl}}\subseteq B_r$. An external acknowledgment establishes neither control nor verified removal from that party's state within the requested scope.

Within this boundary, removal must account for both the target-bearing state and the routes through which its influence can persist or return. The operational closure $\mathrm{Cl}_t(r)$ identifies the required direct, derived, and restorative state (Equation~\ref{eq:operational-closure}). The associated finite path inventory $\mathcal{P}_r$ records routes through which this state can reveal or reconstruct the target, or cause actions based on it. Known but inaccessible paths remain untested obligations, while unknown lineage is disclosed separately. These state and path descriptions define what evaluation must cover. For evolving state, a counterfactual reference may be constructed by replaying history without the target event, holding other inputs fixed where feasible and specifying which events can be replayed and how the resulting histories are compared.

\subsection{Operational Terms and Evidence-Supported Claims}
\label{sec:claim-definitions}

The state and control boundaries above help distinguish operations that produce different outcomes. \emph{Data deletion} removes a source record, whereas \emph{state deletion} removes a context, index, cache, or memory item. Neither operation alone establishes that the target's influence has been removed from dependent state. \emph{Behavioral suppression} blocks an output while leaving that influence potentially recoverable~\cite{nips/CooperEtAl25}. \emph{Knowledge editing} replaces an association; it supports a removal claim only to the extent that residual prior influence is evaluated~\cite{aaai/LiWSKQCWL26}. \emph{Parametric unlearning} seeks target removal by changing parameters or knowledge-bearing adapters, but a parameter change alone does not establish success. \emph{System-level unlearning} coordinates interventions across the requested components and their dependent information paths, then verifies that removal persists through subsequent interactions and system updates.

An empirical removal claim requires low residual target influence, sufficient retained utility, and limited non-target change over the required paths and test conditions. Let $t_u$ denote intervention completion time, $\mathcal{A}_r$ the evaluated access or attack conditions, and $\mathcal{H}_r\subseteq[0,H_r]$ the elapsed observation times, including $0$ for immediate evaluation. State $\mathcal{S}_{t_u+h}^{u}$ follows the declared evolution policy, allowing later tests to assess removal after updates and interactions.

Let $V_{p,a,h}$ measure residual influence on path $p$ under condition $a$ at elapsed time $h$, with lower values indicating less influence. Policy $\tau_r$ distinguishes forbidden influence from permitted reuse. Scores $Q_h$ and $D_h$ measure retained utility and non-target deviation at the deployed state. Each needs an evaluator, reference, sampling budget, and uncertainty estimate. Thresholds $\epsilon_p$, $\eta_r$, and $\kappa_r$ and rules for accounting for uncertainty are declared before judging success. Section~\ref{sec:system-evaluation} explains test design, reference comparisons, and the separate evaluation of attacks that change system state.

A pass requires valid, completed tests and successful inspections of source deletion, replica invalidation, and other required changes. Failed inspections rule out a pass; missing checks leave it unsupported. With these prerequisites:
\begin{equation}
\mathsf{Pass}_{\mathrm{eval}}(r)=1
\quad\Longleftrightarrow\quad
V_{p,a,h}\leq\epsilon_p,\qquad Q_h\geq\eta_r,\qquad D_h\leq\kappa_r.
\label{eq:claim-contract}
\end{equation}
The conditions hold for every $p\in\mathcal{P}_r$, $a\in\mathcal{A}_r$, and $h\in\mathcal{H}_r$, with all three sets nonempty. For coverage reporting, $z_p=1$ records completed tests for path $p$, regardless of outcome. Missing tests leave claims unsupported.

The six evidence dimensions describe different parts of this assessment. \emph{Efficacy} concerns residual target influence, \emph{specificity} non-target deviation, and \emph{utility} retained performance. \emph{Robustness} examines different access or attack conditions, \emph{persistence} checks later observations, and \emph{completeness} concerns coverage of the required state and paths. Together they form the evidence profile $\mathcal{E}_r$, which records tests, results, and assumptions for each dimension rather than a single aggregate score. Passing the protocol supports removal only for the evaluated paths, conditions, and times. It neither establishes coverage of unknown paths or intervals between observations nor supplies a formal guarantee; a certificate requires a proof whose assumptions cover the claimed model, data, and evaluation domain.

Claim boundaries identify the state covered by the evidence. A \emph{model-level claim} covers parameters, representations, adapters, and checkpoint behavior; a \emph{system-level claim} includes required external or persistent state and end-to-end paths. A \emph{boundary-limited claim} identifies required state beyond modification or verification. These labels can overlap. The requested boundary $B_r$ stays fixed; excluding a failed or untested component cannot turn the result into a pass.

\subsection{Authorization and Security Conditions}
\label{sec:unlearning-attack-surface}

The preceding decision rule assumes that the requested change is authorized and that its evaluation covers relevant harm. These assumptions matter because the removal process can itself expose information or damage retained behavior. Checkpoint differencing can reveal removed features~\cite{sp/HuWDX24}, malicious clients can exploit excessive removal~\cite{ndss/HuWCZSHZM24}, and prepared data or requests can activate backdoors or remove refusal behavior~\cite{usenix/HuangMZ24,usenix/Song25}. Authentication and target validation establish which changes are permitted; controls on checkpoint access and post-update tests address risks created by those changes. These conditions qualify the removal claim rather than adding another evidence dimension.

%% file: sections/04_lifecycle_taxonomy.tex
\section{A Unified Framework for LLM Unlearning}
\label{sec:framework}

The five-layer framework applies Section~\ref{sec:foundations} to study comparisons. It connects the request to system scope, information locus, intervention, and evidence-backed claim. Each step constrains the next: the target determines which state matters, while changes and tests determine the supported outcome. Untested components remain outside the result.

\subsection{How the Five Layers Guide Comparison}
\label{sec:framework-comparison}

\emph{Request semantics} records the target, its granularity, preservation requirements, and whether requests are isolated or repeated. \emph{System scope} establishes the components and access available in the evaluated setting. \emph{Information locus} then identifies the specific parameters, records, or derived state that may carry the target within that scope. \emph{Intervention} describes which of these locations the method changes and how. Finally, the \emph{evidence-backed claim} links the reported outcome to its tests, references, and remaining limits. Unreported tests and changes remain unknown.

ToolDelete illustrates all five layers~\cite{icml/ChengA25}. The \emph{request} targets learned tool capabilities while preserving other tools and general tasks. The \emph{scope} is a tool-trained checkpoint; the \emph{locus} is its parameters. The \emph{intervention} updates them using tool-free responses and retention objectives. The \emph{claim} concerns reduced deleted-tool performance, supported by retained-task and membership tests. It does not cover information left by earlier tool calls in external files or services. Following the layers thus separates the requested capability change from broader, untested state deletion.

The same layers distinguish NPO's parameter updates against forget examples~\cite{colm/ZhangLBM24} from the memory audit's record deletion and extraction tests~\cite{corr/Lei2606Agents}. Reduced answer probability and deleted entries concern different state and references. Sections~\ref{sec:standalone} and~\ref{sec:augmented} compare these mechanisms within their evaluated boundaries, without ranking incompatible scores.

\begin{table*}[!ht]
\caption{How the five layers organize the review. Layers can apply across several sections.}
\label{tab:framework-roadmap}
\centering
\small
\setlength{\tabcolsep}{4pt}
\renewcommand{\arraystretch}{1.08}
\arrayrulecolor{stageblue!65}
\begin{tabularx}{\textwidth}{L{.22\textwidth}L{.47\textwidth}Y}
\toprule
\rowcolor{stageblue!16}
\textbf{Layer} & \textbf{Comparison question} & \textbf{Main discussion} \\
\midrule
\cellcolor{stageblue!8}Request semantics & What is removed, and what must remain? & Sections~\ref{sec:standalone-benchmarks},~\ref{sec:applications} \\
\rowcolor{stageblue!4}
\cellcolor{stageblue!8}System scope & Which components and access are included? & Sections~\ref{sec:standalone},~\ref{sec:augmented} \\
\cellcolor{stageblue!8}Information locus & Where can the target reside or affect behavior? & Sections~\ref{sec:standalone-methods},~\ref{sec:augmented} \\
\rowcolor{stageblue!4}
\cellcolor{stageblue!8}Intervention & What changes, and what does the change require? & Sections~\ref{sec:standalone-methods},~\ref{sec:augmented} \\
\cellcolor{stageblue!8}Evidence-backed claim & What do the tests establish, and what is untested? & Sections~\ref{sec:standalone-evaluation},~\ref{sec:system-evaluation},~\ref{sec:trends} \\
\bottomrule
\end{tabularx}
\end{table*}

Table~\ref{tab:framework-roadmap} connects the comparison questions to later chapters. The layers organize study analysis; the lifecycle identifies when removal operations and checks occur; and the evidence dimensions organize their evaluation. These views serve different purposes, while preserving each study's evaluated setting and the limits of its reported evidence.

\subsection{System Scope and Expanding Information Loci}

Figure~\ref{fig:evolution} compares four system scopes. Standalone LLM studies address parameters, representations, or adapters; augmented systems add context, retrieval stores, indexes, and caches. Agents introduce persistent memory, plans, tool use, and trajectories, while multi-agent settings add messages, shared state, and independently managed replicas. These scopes do not prescribe an architecture: agents may lack retrieval, and peers may share only selected information.

Scope changes the intervention and its claim. Parameter updates address model state; source deletion must reach required retrieval and cache entries. Agents may require revised plans or repaired tool effects, and distributed removal requires coordination with other owners. Information locus identifies affected state; system scope identifies control and inspection limits. Sections~\ref{sec:standalone} and~\ref{sec:augmented} compare access, interventions, outcomes, and limits at these locations.

\subsection{The Seven-Stage Lifecycle}

\begin{figure*}[!ht]
  \centering
  \input{figures/lifecycle_agentic}
  \caption{Seven-stage unlearning lifecycle with inputs, required outputs, and failure modes.}
  \label{fig:lifecycle}
\end{figure*}

Figure~\ref{fig:lifecycle} organizes execution into seven stages. \emph{Request} specifies the authorized target and preservation policy; \emph{Scope} identifies required components and control limits; and \emph{Localize} traces target-bearing state and its dependencies. \emph{Execute} changes that state, while \emph{Propagate} updates affected derivatives and copies. \emph{Verify} tests the required outcomes, and \emph{Maintain} repeats checks after later changes. The stages describe obligations where applicable: a standalone checkpoint may have no external replica requiring propagation, whereas a RAG request can require cache invalidation and tests after index rebuilding. They identify when the state changes and tests recorded by the five analytical layers occur.

\subsection{Categories and Comparison Conditions}

\begin{figure*}[!ht]
  \centering
  \input{figures/taxonomy_agentic}
  \caption{Five-layer taxonomy for comparing studies. Categories can combine; system scope is distinct from claim boundary.}
  \label{fig:taxonomy-framework}
\end{figure*}

Figure~\ref{fig:taxonomy-framework} supplies categories for the comparison. Granularity distinguishes records, entities, concepts, and domains; request regimes distinguish individual, batch, sequential, and continual processing. These choices are independent of propagation and multi-owner authority, which can arise under any regime. Loci and interventions can combine, while the claim records tested outcomes and unresolved dependencies. Access, cost, guarantee assumptions, and application objectives provide additional comparison conditions rather than evidence dimensions. Comparisons account for data, references, preparation budgets, and evaluated access conditions before attributing results to interventions~\cite{nips/DornaMZMKLM25}.

\subsection{Interpreting System Claims and Evidence Maturity}

A retrieval request illustrates why scope and evidence must remain separate. Removing a personal fact may require changing its source, summaries, and caches. Source deletion addresses one locus; checking derivatives and testing retrieval supports a wider claim. An inaccessible required replica remains unresolved within the original scope.

\smallskip \noindent \textbf{Evidence Maturity.} Figure~\ref{fig:evolution} summarizes the breadth and integration of available tests. Substantial evidence spans benchmarks, method comparisons, and recovery tests; fragmented evidence covers individual parts without integrated tests; early evidence consists of limited tests for particular targets. Open assurance denotes requirements for which integrated evidence remains unavailable. These descriptions concern evidence coverage rather than architectural complexity or ratings of individual papers. Comparative removal studies and memory-management evaluations illustrate why both test breadth and the outcome tested matter~\cite{nips/DornaMZMKLM25,acl/XiaoMCYZYCC26,acl/LeeLX26,acl/XiongLXHLTLX26}. Section~\ref{sec:trends} applies these descriptions.

%% file: figures/lifecycle_agentic.tex
\includegraphics[width=\textwidth]{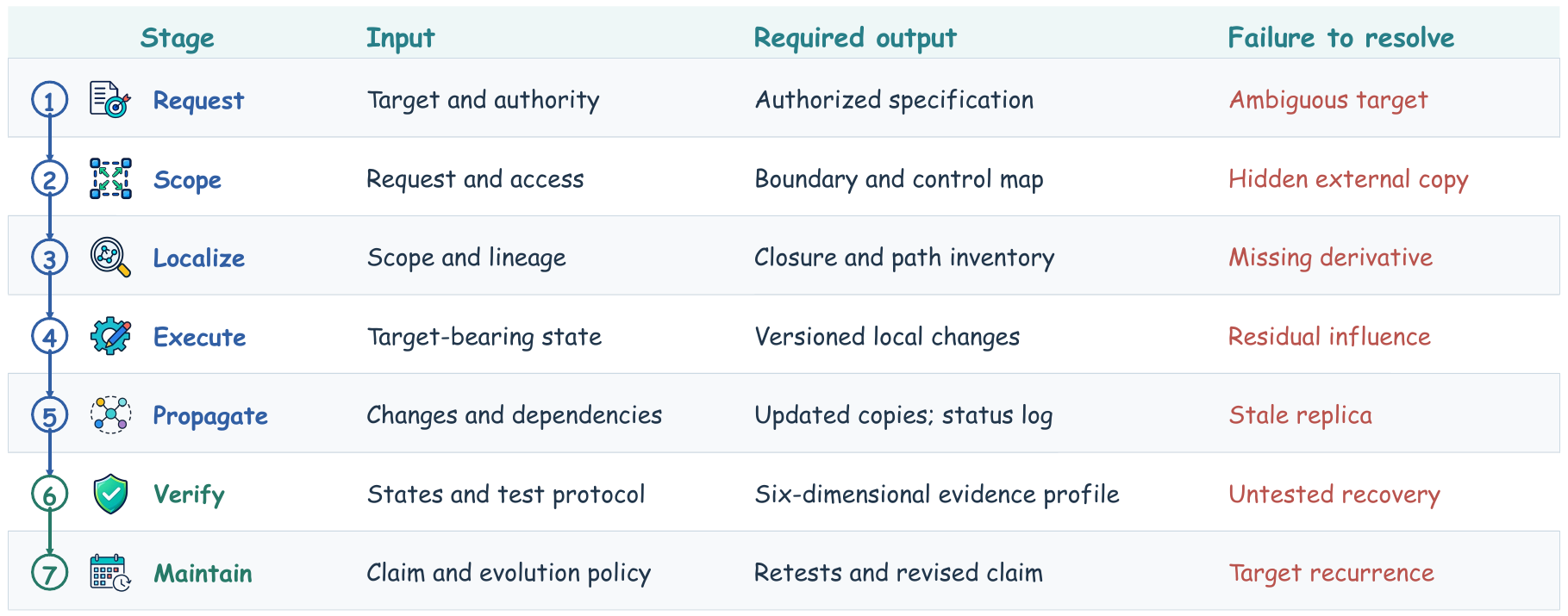}

%% file: figures/taxonomy_agentic.tex
\includegraphics[width=\textwidth]{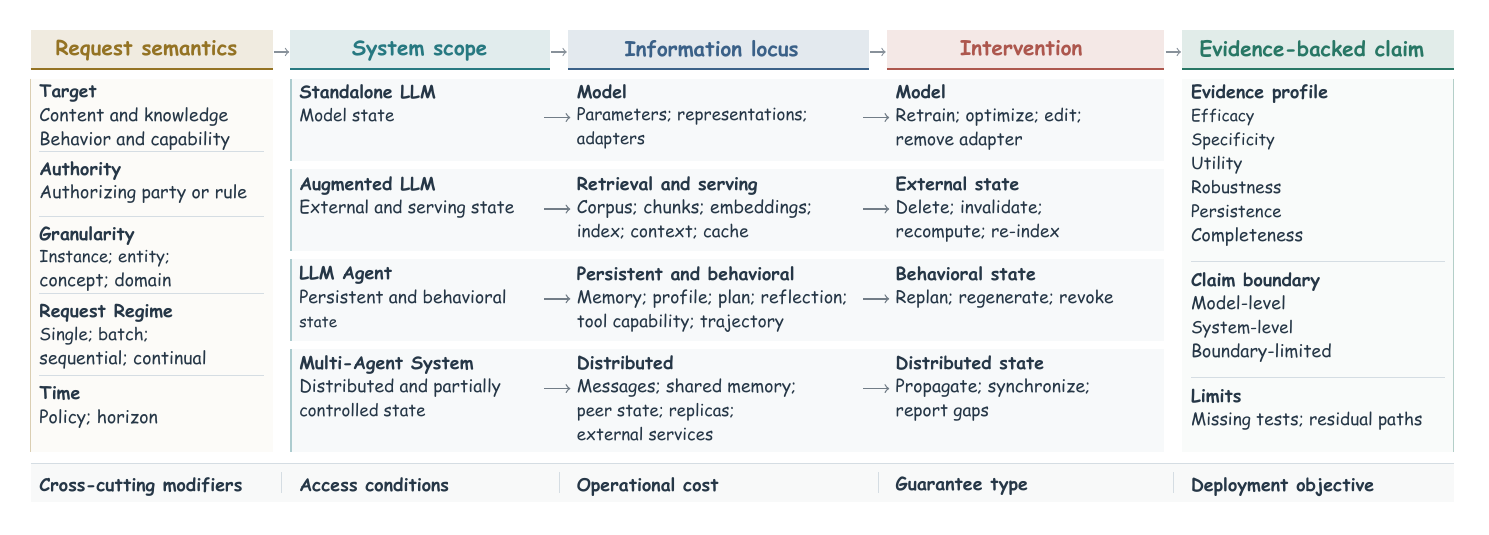}

%% file: sections/05_standalone_llms.tex
\section{LLM Unlearning: Benchmarks, Methods, and Evidence}\label{sec:standalone}
We apply the five-layer comparison first to standalone LLMs, fixing the system scope to the model and its inference interface. Section~\ref{sec:standalone-benchmarks} examines targets and request regimes; Section~\ref{sec:standalone-methods} compares changed state, mechanisms, and access requirements; and Section~\ref{sec:standalone-evaluation} assesses what the reported tests support. This order connects benchmark choices to method behavior and then to the limits of the resulting claim. Section~\ref{sec:augmented} extends this analysis beyond the model.

\subsection{Benchmarks: What Should Be Forgotten?}\label{sec:standalone-benchmarks}
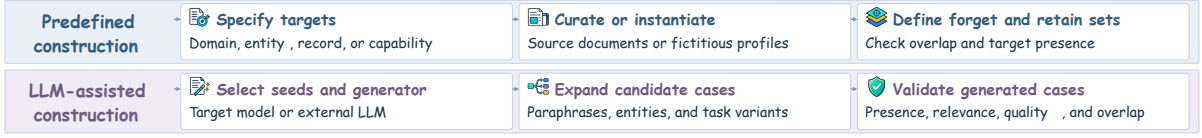
\begin{figure*}[!t]
    \centering
    \input{figures/benchmark_construction_agentic}
    \caption{Predefined and LLM-assisted benchmark construction, from target specification to forget/retain set validation.}
    \label{fig:benchmark}
\end{figure*}

Benchmark design determines how the target $T_r$, granularity $g_r$, and retain boundary become test cases. This makes construction a bottleneck in evaluation. Efficacy is distorted when targets do not match model knowledge~\cite{satml/Thaker0KMWS25}, and privacy rankings vary with data, model, method, and attack configuration~\cite{aaai/QianZLH26}. A credible suite therefore checks target presence and related retained knowledge and specifies the request regime. Figure~\ref{fig:benchmark} contrasts two construction routes, predefined and LLM-assisted cases. Both require checks of target presence, forget-retain overlap, and case quality.

\begin{table*}[!t]
\caption{Representative benchmarks and forget-set construction methods. SFT: supervised fine-tuning; $\dagger$: preprint.}
\label{tab:request_types}
\centering
\scriptsize
\setlength{\tabcolsep}{4pt}
\renewcommand{\arraystretch}{1.10}
\arrayrulecolor{stageblue!65}
\begin{tabularx}{\textwidth}{@{}L{0.22\textwidth}L{0.22\textwidth}L{0.20\textwidth}Y@{}}
\toprule
\rowcolor{stageblue!16}
\tablecolorleft{stageblue!16}\textbf{Benchmark / approach} & \textbf{Construction} & \textbf{Target} &\tablecolorright{stageblue!16} \textbf{Evaluation focus} \\
\midrule
\tablecolorleft{stageblue!8}WHP$^{\dagger}$~\cite{corr/eldan23} & Fixed corpus & Copyright domain & Recall and reconstruction \\
\rowcolor{stageblue!4}
\tablecolorleft{stageblue!8}TOFU~\cite{colm/Maini24} & Fictitious profiles & Synthetic author facts &\tablecolorright{stageblue!4} Forget-utility balance \\
\tablecolorleft{stageblue!8}RWKU~\cite{nips/JinCWHYL00024} & Real entities & Entity knowledge & Related-fact specificity \\
\rowcolor{stageblue!4}
\tablecolorleft{stageblue!8}WMDP~\cite{icml/LiPGYBGLDGMHLJL24} & Expert-authored QA & Hazard proxy &\tablecolorright{stageblue!4} Knowledge reduction; benign utility \\
\tablecolorleft{stageblue!8}MUSE~\cite{iclr/shi24} & Books and news & Copyright, privacy & Multi-axis efficacy and privacy \\
\rowcolor{stageblue!4}
\tablecolorleft{stageblue!8}KnowUnDo~\cite{emnlp/TianL0LWS0C024} & Forget-retain pairs & Fact / entity &\tablecolorright{stageblue!4} Over-unlearning \\
\tablecolorleft{stageblue!8}WPU~\cite{emnlp/LiuZJC24} & Wikipedia biographies & Person-specific facts & Fabrication; related-fact retention \\
\rowcolor{stageblue!4}
\tablecolorleft{stageblue!8}LUME~\cite{emnlp/RamakrishnaWJCBVCHG25LUME} & Three-task suite & Document / entity &\tablecolorright{stageblue!4} Cross-task method comparison \\
\tablecolorleft{stageblue!8}SemEval Task 4~\cite{semeval/RamakrishnaWJCBVCHG25} & Shared task & Privacy, copyright & Common tracks and models \\
\rowcolor{stageblue!4}
\tablecolorleft{stageblue!8}DUAL~\cite{acl/BorisiukSPT26} & Wikidata triplets & Fact / salience &\tablecolorright{stageblue!4} Pretraining versus SFT origin \\
\tablecolorleft{stageblue!8}DUSK~\cite{acl/JeungYHKSHYN26} & Overlapping documents & Shared knowledge & Forget-specific versus shared content \\
\rowcolor{stageblue!4}
\tablecolorleft{stageblue!8}PCH$^{\dagger}$~\cite{corr/xu2026pch} & Request stream & Mixed targets &\tablecolorright{stageblue!4} Continual drift and interactions \\
\tablecolorleft{stageblue!8}ORT~\cite{corr/Ye2506Form} & Expression variants & Concept / capability & Form-dependent bias \\
\midrule
\tablecolorleft{stageblue!8}Textbook~\cite{colm/Zhu25} & LLM-generated textbooks & Domain knowledge & No expert-curated forget set \\
\rowcolor{stageblue!4}
\tablecolorleft{stageblue!8}BiForget~\cite{acl/xu26} & Target model & Domain / instance &\tablecolorright{stageblue!4} Relevance, diversity, quality, cost \\
\bottomrule
\end{tabularx}
\end{table*}

Table~\ref{tab:request_types} separates targets from construction approaches. WMDP measures hazardous-knowledge QA, not harmful action. DUAL separates pretraining from SFT, DUSK exposes shared knowledge, PCH models interacting requests, and ORT varies expression~\cite{acl/BorisiukSPT26,acl/JeungYHKSHYN26,corr/xu2026pch,corr/Ye2506Form}. KnowUnDo and WPU stress collateral damage and fabrication~\cite{emnlp/TianL0LWS0C024,emnlp/LiuZJC24}; coreset effects show that few tokens can drive apparent success~\cite{corr/Pal2504Coreset}. Textbook and BiForget generate forget sets, but target presence, overlap, and retained knowledge still require model-specific validation of prior influence~\cite{colm/Zhu25,acl/xu26}.

\smallskip \noindent \textbf{Request Regimes.}
A \emph{single} request sets target and retain boundaries; batches group targets; sequential or continual requests form streams. Across a request stream, drift may restore previously removed influence, harm utility, or raise audit cost~\cite{corr/Barez25}. PCH tests request interactions; DUSK and ORT expose overlap and expression coupling~\cite{corr/xu2026pch,acl/JeungYHKSHYN26,corr/Ye2506Form}. Comparisons must therefore state the request units, their order and overlap, and the evaluation horizons.

\subsection{Methods: How Can a Model Forget?}\label{sec:standalone-methods}






For a specified request, method choice depends on access, data, and the required outcome. We first compare interventions that change inference context, output probabilities, internal representations, or selected parameter directions. We then examine low-rank updates, adversarial training, and continual processing as compatible designs. Table~\ref{tab:unlearning_taxonomy} summarizes representative choices; the discussion links their requirements to reported results and practical limitations.

\begin{table*}[!t]
\caption{Representative mechanisms and compatible update designs, with requirements and principal trade-offs.}
\label{tab:unlearning_taxonomy}
\centering
\scriptsize
\setlength{\tabcolsep}{4pt}
\renewcommand{\arraystretch}{1.10}
\arrayrulecolor{stageblue!65}
\begin{tabularx}{\textwidth}{@{}L{0.19\textwidth}L{0.28\textwidth}L{0.23\textwidth}Y@{}}
\toprule
\rowcolor{stageblue!16}
\tablecolorleft{stageblue!16}\textbf{Approach} & \textbf{Required access / data} & \textbf{Changed state / operation} &\tablecolorright{stageblue!16} \textbf{Main limitation / trade-off} \\
\midrule
\multicolumn{4}{c}{\textit{Representative intervention mechanisms}} \\
\tablecolorleft{stageblue!8}ICUL~\cite{icml/PawelczykNL24} & Queries; forget + retain & Inference context & Prompt dependence \\
\rowcolor{stageblue!4}
\tablecolorleft{stageblue!8}NPO~\cite{colm/ZhangLBM24} & Weights; forget; reference & Weights: output loss &\tablecolorright{stageblue!4} Retain-loss balance \\
\tablecolorleft{stageblue!8}RMU~\cite{icml/LiPGYBGLDGMHLJL24} & Weights; forget + retain & Weights: activation loss & Retain-set coverage \\
\rowcolor{stageblue!4}
\tablecolorleft{stageblue!8}PISCES~\cite{emnlp/gur2025} & Weights; SAE; validation & MLP parameter directions &\tablecolorright{stageblue!4} Feature coverage \\
\midrule
\multicolumn{4}{c}{\textit{Compatible designs for updates, recovery, and repeated requests}} \\
\tablecolorleft{stageblue!8}ReGLU~\cite{acl/XiaoMCYZYCC26} & Weights; forget + retain & Low-rank updates & Subspace estimation \\
\rowcolor{stageblue!4}
\tablecolorleft{stageblue!8}LAU~\cite{aaai/YuanJCCLZ25} & Weights; model activations & Latent attack and defense &\tablecolorright{stageblue!4} Attack-training cost \\
\tablecolorleft{stageblue!8}$O^3$~\cite{iclr/gao25} & Weights; forget only & LoRA; input detector & Detector dependence \\
\bottomrule
\end{tabularx}
\end{table*}

\smallskip \noindent \textbf{Inference-Time Control.} ICUL uses query access and prepends forget examples with altered labels or answers alongside correctly labeled retain examples~\cite{icml/PawelczykNL24}. Its classification experiments reduce confidence on targeted instances while maintaining test accuracy without parameter updates. The trade-off is continued dependence on the supplied context, including its length and composition. This makes ICUL relevant when weights are inaccessible and inference inputs can be controlled; its evaluated effect belongs to that inference configuration. Other controls use routing, auxiliary logits, or learned prompts~\cite{corr/thaker2403,nips/Liu24,nips/JiLZLK0C24,naacl/BhailaVW25,acl/TakashiroKGCIM25,acl/WangGPPYCOLLT26}, with access determined by the modified component.

\smallskip \noindent \textbf{Output-Loss Optimization.} Gradient ascent increases loss on forget examples, but updates can damage general generation. NPO instead uses an adaptive weighting of this gradient relative to the original model: examples receive less weight as their response probability falls~\cite{colm/ZhangLBM24}. It therefore requires parameter access, forget examples, and reference probabilities; retain regularization additionally requires retained data. On TOFU with Llama-2-7B-Chat, NPO variants improve the reported forget--utility trade-off over gradient-ascent baselines. The retain-loss ablation is particularly informative: increasing its weight initially improves both utility and forget quality, but further increases reduce forget quality. Thus, preservation and removal need not conflict at every setting, and the loss balance matters alongside the chosen objective. Gradient correction and constrained objectives offer related ways to manage this balance~\cite{icml/WangWFLHDH25,nips/Fan25,nips/EntesariHKRF25}.

\smallskip \noindent \textbf{Representation and Parameter Editing.} RMU changes weights to redirect activations on forget data toward a fixed random vector, while matching the original model's activations on retain data~\cite{icml/LiPGYBGLDGMHLJL24}. Unlike output-loss optimization, it specifies the change at an internal layer and requires both datasets and access to model activations. On Zephyr-7B, RMU reduces WMDP-Bio and WMDP-Cyber accuracy while largely preserving the original model's MMLU performance. The preservation term is therefore central to interpreting the result: it protects the sampled benign activations, and its coverage depends on the retain data. Layer choice and steering strength also affect what the update changes, making these settings part of the comparison with methods that optimize output probabilities directly.

PISCES edits concept-related MLP directions using a sparse autoencoder (SAE) and feature selection~\cite{emnlp/gur2025}. On Gemma 2 and Llama 3.1, its concept-erasure experiments report stronger specificity and relearning robustness than the compared methods, with modest efficacy gains. This advantage depends on finding features that isolate the concept: missing features and reconstruction errors limit the edit, and attention parameters remain outside its implementation. Its reported editing-cost advantage also assumes an available SAE and excludes SAE training. PISCES therefore shifts part of the work from repeated optimization to feature discovery and validation. Other localized approaches select tokens, neurons, or subspaces~\cite{emnlp/WuLXDW0X23,emnlp/WanRCGCG25,acl/LeeLX26,emnlp/WenFGWLSGS25,emnlp/WangWLHZ25,iclr/Hang2026clue}; these choices specify where different losses act.

\smallskip \noindent \textbf{Matched Comparisons.} OpenUnlearning shows how evaluation choices affect these conclusions under matched TOFU tests~\cite[Tables~3 and~6]{nips/DornaMZMKLM25}. SimNPO outranks RMU on the harmonic mean of memorization, privacy, and utility; RMU suppresses memorization more but loses utility. GradDiff's departure from the retain reference harms its privacy score despite strong suppression, and excluding privacy reverses the ranking in its favor. This explains why greater target suppression need not identify the preferable method even on the same benchmark. Comparisons must specify the preserved capabilities, privacy criterion, and aggregation rule alongside the intervention and its tuning budget.

\smallskip \noindent \textbf{Parameter-Efficient Updates.} ReGLU illustrates why update size should be separated from the removal objective. It initializes LoRA using forget and retain representations, then constrains updates away from the retain subspace~\cite{acl/XiaoMCYZYCC26}. On WMDP with Zephyr-7B-beta, combining gradient difference with ReGLU lowers mean Bio/Cyber accuracy to 35.1\%, compared with 38.1\% for VILA, while all reported checkpoints preserve at least 95\% of original MMLU utility. This comparison holds the loss fixed and varies the low-rank design. TOFU results also show gains with different losses, although subspace estimation depends on representative data. Its initialization is faster than gradient-based FILA in the reported tests, but covariance estimation and matrix decomposition still incur cost. These results distinguish savings from low-rank updates from the additional work needed to select those updates. Adapter and expert-based approaches make other choices about where updates are stored and which components must remain active~\cite{emnlp/ChenY23,aaai/HuLHZLZ24,iclr/ChaCHL25,icml/TanLCQCG26}.

\smallskip \noindent \textbf{Training Against Recovery.} LAU alternates between learning latent perturbations that recover target knowledge and updating the model to resist them~\cite{aaai/YuanJCCLZ25}. This adds attack optimization to the removal process, targeting recovery that a standard forget loss may overlook. Robustness-aware approaches can therefore accompany parameter or representation updates rather than forming an exclusive mechanism family~\cite{iclr/ZhangWLWTL00W25,icml/FanJZRH025,iclr/Han2026dualspace,iclr/yang2026erase,acl/LiZGWDLWL26}. Their added training work must be considered alongside the evaluated attack access and budget. Comparisons include both training and other recovery conditions, so resistance to a trained attack is not mistaken for broader protection.

\smallskip \noindent \textbf{Sequential and Continual Requests.} SSU identifies content-related weight updates and combines their removal with random-label training to balance copyright-target removal and general language performance over successive requests~\cite{naacl/DouLLDW25}. $O^3$ works without retain data, combining orthogonal LoRA updates with an input detector that controls their activation~\cite{iclr/gao25}. Across three tasks and seven datasets, it reports improved utility retention over its baselines. That benefit depends on the detector and adapter configuration, which also adds inference computation and storage. Together, these examples separate two design choices: how to limit interference between requests and how to preserve behavior without retained examples. Other stream-oriented methods localize, filter, or gate updates~\cite{icml/wuerkaixi2025adaptive,corr/xu2026pch,acl/0019DTLZLGW0H25,corr/li2025}. Comparisons fix request order and overlap, retest earlier targets, and record total storage and audit costs.

These comparisons guide intervention choice under specific conditions. Query-only access favors inference controls; direct model updates require suitable data, reference information, or identifiable features. Low-rank and recovery-aware designs then address resource and attack constraints, while repeated requests introduce cumulative effects. Section~\ref{sec:standalone-evaluation} examines the metrics needed to assess these choices without hiding their trade-offs in a single score.

\subsection{Model-Level Evidence: What Does Evaluation Establish?}\label{sec:standalone-evaluation}

\begin{figure*}[!t]
    \centering
    \resizebox{0.85\textwidth}{!}{%
        \input{figures/evaluation_agentic}%
    }
    \caption{Evaluation of standalone LLMs across six evidence dimensions.}
    \label{fig:evaluations}
\end{figure*}
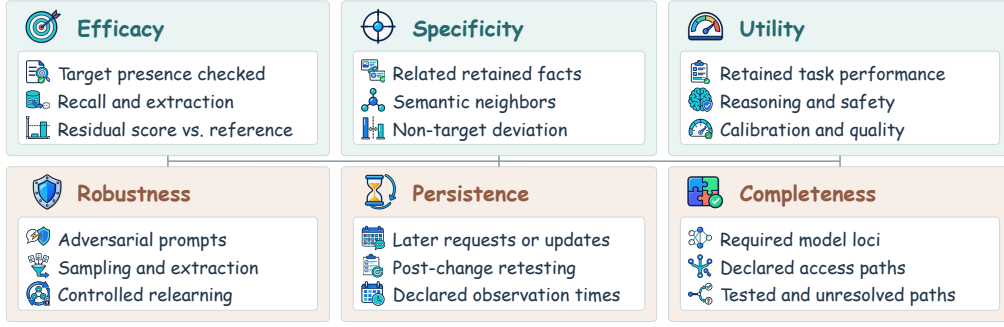

Figure~\ref{fig:evaluations} applies the evidence dimensions defined in Section~\ref{sec:claim-definitions} to model-level tests. The following comparisons examine what particular metrics reveal and why evaluation choices can change the interpretation of a model update.

\smallskip \noindent \textbf{Efficacy, Specificity, and Utility.} Accuracy, likelihood, perplexity, and distribution tests capture distinct behaviors~\cite{acl/YaoCDNWCY24,colm/Maini24}. TOFU's Truth Ratio contrasts the likelihoods of perturbed answers and paraphrased correct answers. Related-fact and semantic-neighbor probes complement broad retained-task scores by testing whether an update damages knowledge close to the target. OpenUnlearning standardizes comparisons and meta-evaluates metrics~\cite{nips/DornaMZMKLM25}; FPE samples semantic variants, UNCD decomposes hazardous capability, and REMIND probes local geometry~\cite{emnlp/ChengHW25,acl/LangGHZZYSZ25,acl/CohenNM26}. Aggregate scores can hide generative and recovery failures despite reduced average target accuracy~\cite{corr/Fan2510Microscope}.

\smallskip \noindent \textbf{Recovery, Privacy, and Persistence.} Prompting, representation probes, relearning, membership attacks, quantization, and probabilistic decoding test different recovery channels~\cite{corr/Lynch2402,iclr/PatilHB24,emnlp/ToL25,iclr/0001FWS25,iclr/ZhangWLWTL00W25,icml/ReisizadehRCPLH26}. Checkpoint differencing and textual audits expose residual privacy signals~\cite{icml/DuWZPHR25,nips/WuPLW25}, while soft-token attacks require negative controls to avoid false positives~\cite{emnlp/ChenSXH25}. RULI and privacy benchmarking show that efficacy and privacy must be measured jointly at sample level~\cite{usenix/Naderloui25,aaai/QianZLH26}; MNEME and PreUnlearn probe collateral effects beyond predefined retained-task suites~\cite{emnlp/KassemSRF25,corr/Su2606PreUnlearn}.

\smallskip \noindent \textbf{Operational Cost.} Model comparisons also depend on the resources required to construct the forget set and apply the update. Reports measure compute, memory, runtime, and construction cost~\cite{acl/YaoCDNWCY24,emnlp/xu25}; repeated requests add storage, verification, and consolidation demands that affect service throughput~\cite{acl/xu26}. These measurements help explain whether an approach remains practical as request volume grows. Section~\ref{sec:system-evaluation} integrates costs into evaluation reporting.

The next section follows target information into retrieval stores, caches, and memory. These components can return information after a model update, shifting the analysis from checkpoint behavior to the dependencies supplying its inputs. Equation~\ref{eq:operational-closure} identifies the additional state and connects source changes to retrieval and serving checks.

\begin{tcolorbox}[colback=softblue,colframe=stageblue!70,boxrule=0.8pt,arc=3pt]
\textbf{Takeaway:} Model-level evidence depends on matching benchmark targets, intervention scope, and evaluation conditions. Target-presence checks and retained-data boundaries anchor comparisons; recovery and collateral-effect tests reveal failures hidden by aggregate scores. For repeated requests, retesting earlier targets and recording cumulative costs show how removal and retained performance change as updates accumulate.
\end{tcolorbox}

%% file: figures/benchmark_construction_agentic.tex
\includegraphics[width=\textwidth]{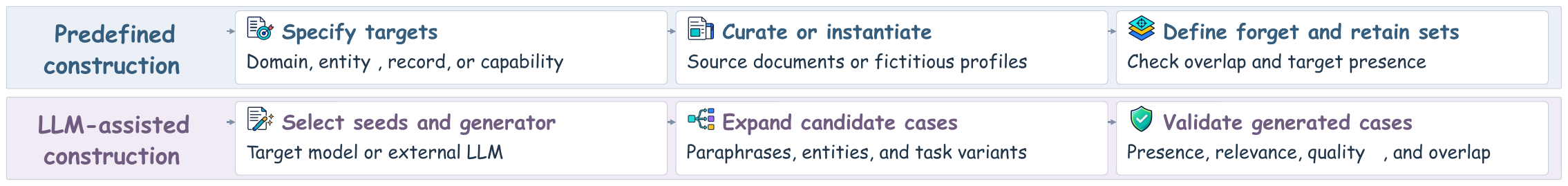}

%% file: figures/evaluation_agentic.tex
\includegraphics[width=\textwidth]{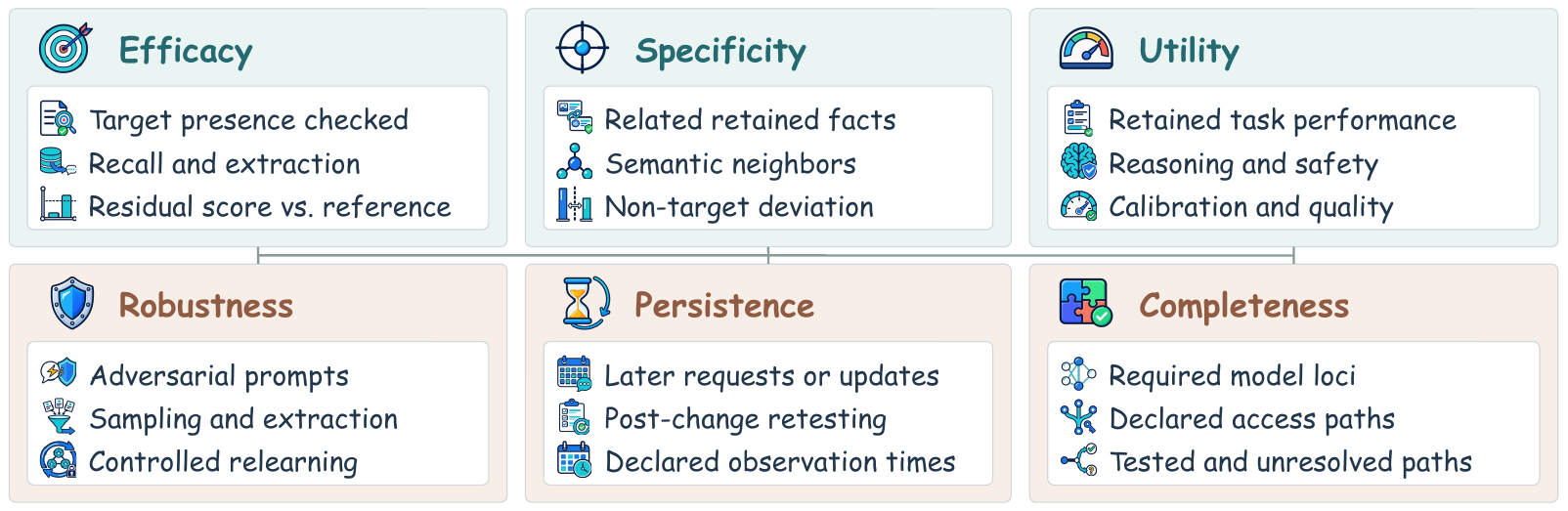}

%% file: sections/06_augmented_llms.tex
\section{Extending Unlearning Beyond the Model}
\label{sec:augmented}

A model update can leave target information available through retrieval, serving, or agent state. We retain the five-layer comparison from Section~\ref{sec:standalone}, but expand the system scope and information locations under review. For each setting, we distinguish the removal target, changed state, intervention, reported tests, and claim boundary. This distinguishes tested removal from supporting mechanisms and untested requirements for distributed coordination.

\subsection{Augmented LLMs: Retrieval, Context, and Serving State}

Augmented LLMs combine a model with context, retrieval, and serving state~\cite{nips/LewisPPPKGKLYR020,corr/Gao2312RAGSurvey,sigir/ShiX0Z00025}. The chain corpus $\rightarrow$ chunks $\rightarrow$ embeddings $\rightarrow$ index $\rightarrow$ cache is a dependency template, not a fixed architecture: caches may store retrieved items or generated responses, and context can receive direct user input. The request tuple is unchanged: $B_r$ covers the affected source, derivative, and runtime state, while $B_r^{\mathrm{ctrl}}$ records its directly controllable and inspectable subset.
Deletion, invalidation, recomputation, and re-indexing must follow these dependencies, while model updates address any in-scope parametric influence. Evaluation then probes retrieval, contextual injection, and restoration of the target. The ability to edit individual components alone does not establish removal across all required state and retrieval paths.

\smallskip \noindent \textbf{Externalization and Residual Influence.} Deleting external state does not itself remove parametric knowledge. More generally, \emph{split forgetting} denotes local success while another required path still exposes, reconstructs, or acts on the target.
Existing evidence illustrates why the claim boundary matters. RAG-based unlearning retrieves request-specific external knowledge to steer responses without changing model parameters~\cite{corr/Wang2410RAG}; exact in-context unlearning applies to a designed adaptation state~\cite{icml/MuresanuTZP25}; and Limited Memory Language Models externalize factual values by construction~\cite{iclr/ZhaoZBLZNGWAS26}. These results support retrieval-conditioned, adaptation-state, or architecture-specific claims rather than unrestricted parametric erasure. Context-aware unlearning separates training recall from authorized contextual use~\cite{icml/PengAGBHWH26}.

\smallskip \noindent \textbf{Interventions and Required Checks.} Table~\ref{tab:augmented-loci} pairs candidate interventions with the local checks they require. These checks are obligations, not guarantees: restricting access differs from deletion, and withdrawn adapters may remain recoverable. A system claim must combine checks across all required derivatives and serving paths.

\begin{table*}[!ht]
\caption{Candidate interventions and local checks by information locus in augmented LLMs. Repr.: representations.}
\label{tab:augmented-loci}
\centering
\scriptsize
\setlength{\tabcolsep}{4pt}
\renewcommand{\arraystretch}{1.08}
\arrayrulecolor{stageblue!65}
\begin{tabularx}{\textwidth}{@{}L{0.17\textwidth}L{0.24\textwidth}L{0.23\textwidth}Y@{}}
\toprule
\rowcolor{stageblue!16}
\tablecolorleft{stageblue!16}\textbf{Locus} & \textbf{In-scope state} & \textbf{Intervention} &\tablecolorright{stageblue!16} \textbf{Required local check} \\
\midrule
\tablecolorleft{stageblue!8}Model state & Parameters; repr.; adapters & Update; edit; remove adapter & Residual recall within threshold \\
\rowcolor{stageblue!4}
\tablecolorleft{stageblue!8}Corpus & Documents; source versions & Delete; redact; restrict access &\tablecolorright{stageblue!4} Absence or disclosed restriction \\
\tablecolorleft{stageblue!8}Chunks & Windows; overlaps; summaries & Delete; recompute & Target absent from required derivatives \\
\rowcolor{stageblue!4}
\tablecolorleft{stageblue!8}Embeddings & Vectors; source mappings & Invalidate; recompute &\tablecolorright{stageblue!4} Target vectors no longer served \\
\tablecolorleft{stageblue!8}Index & Keys; postings; replicas & Re-index; synchronize & Target absent in retrieval tests \\
\rowcolor{stageblue!4}
\tablecolorleft{stageblue!8}Context & Prompts; examples; tool returns & Remove; recompute &\tablecolorright{stageblue!4} Replay excludes targeted content \\
\tablecolorleft{stageblue!8}Cache & Keys; responses; replicas & Invalidate; flush & No target in warm-cache probes \\
\bottomrule
\end{tabularx}
\end{table*}

\smallskip \noindent \textbf{Corpora, Chunks, Embeddings, and Indexes.} Let $\mathrm{Seed}_t(T_r;B_r)$ collect target-bearing state within the boundary at request granularity. Closure specifies required state, not an algorithm that discovers all dependencies:
\begin{equation}
\mathrm{Cl}_t(r)=\mathrm{Seed}_t(T_r;B_r)
\cup\mathrm{Desc}_t(\mathrm{Seed}_t(T_r;B_r);B_r)
\cup\mathrm{Restore}_t(T_r;B_r).
\label{eq:operational-closure}
\end{equation}
Here $\mathrm{Desc}_t$ contains required transitive derivatives. $\mathrm{Restore}_t$ contains in-scope state able to restore the target or derivatives under $\tau_r$, transitively expanded over restoring dependencies and required derivatives. Restoration beyond $B_r$ is a disclosed limit. Independent correlated facts enter the closure only when the request includes them. Missing lineage prevents a claim that this inventory is complete.
An old address may pass from a record into a profile summary and cached response. Deleting the record while its address remains cached illustrates split forgetting. The closure must also include snapshots that can restore the address. Inspection, retrieval, and replay then test whether the coordinated intervention has addressed these routes. Incomplete provenance and independently controlled services limit claims.

Knowledge-base editing provides supporting evidence. ERASE maintains consistency as entries are deleted or rewritten~\cite{naacl/LiLRZNA25}, while temporal RAG studies show that updates may require version-aware organization~\cite{acl/LiuCL26}. Neither alone establishes machine unlearning. Verification must cover aliases, paraphrases, overlapping chunks, semantic neighbors, stale snapshots, and re-ingestion; embedding models such as SimCSE are probes, not unlearning methods~\cite{emnlp/GaoYC21}. Likewise, prospective privacy~\cite{ccs/DuYC0H023,www/DuYCS23}, source deletion, and parametric forgetting are distinct guarantees.

\smallskip \noindent \textbf{Context and Caches.} Removing content from the current context does not address its earlier derivatives. Retroactive removal must also cover in-scope logs, summaries, caches, training queues, and tool effects. Prompt-mediated controls~\cite{icml/PawelczykNL24,acl/TakashiroKGCIM25} affect inference, not earlier stored descendants. Derived semantic-cache entries and replicas need invalidation tied to provenance and tests of semantic neighbors for potentially surviving target information.

\smallskip \noindent \textbf{Composing Interventions and Claims.} A system claim composes Table~\ref{tab:augmented-loci}'s postconditions over $\mathrm{Cl}_t(r)$, testing source absence, semantic retrieval, generation with and without retrieval, warm caches, replicas, re-indexing, and parametric reconstruction. Neither RAG accuracy nor local suppression alone establishes this end-to-end composition.

%% file: sections/07_agents_multiagent.tex
\subsection{LLM Agents and Multi-Agent Systems}
\label{sec:agents}





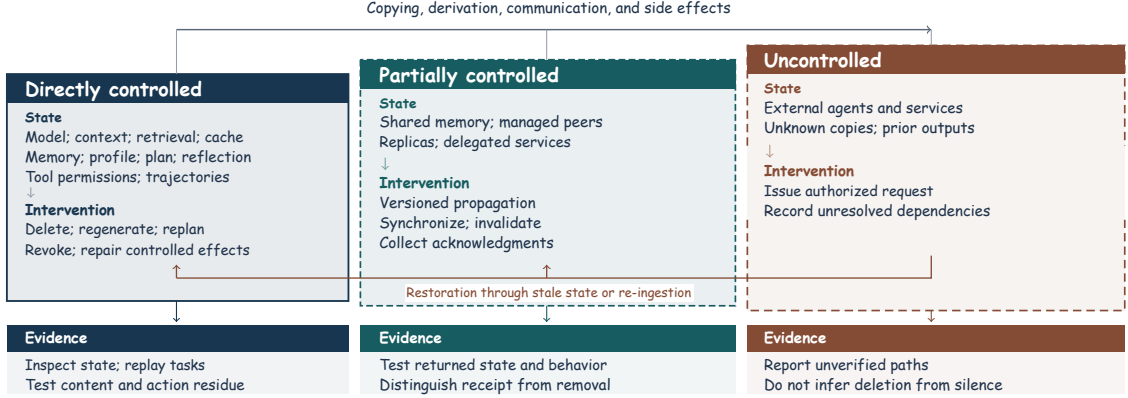
\begin{figure*}[!ht]
  \centering
  \input{figures/agent_anatomy_agentic}
  \caption{State, interventions, and evidence across agentic control boundaries.}
  \label{fig:agent-anatomy}
\end{figure*}

An LLM agent observes, reasons, acts, and updates state. Target information can consequently influence memory, profiles, plans, reflections, tool capabilities, and trajectories, while communication can carry it across ownership boundaries. Figure~\ref{fig:agent-anatomy} distinguishes directly controlled, partially controlled, and uncontrolled state. The early evidence reviewed below tests particular targets and components rather than establishing joint removal across them.

\smallskip \noindent \textbf{Targets and Residual State.} Targets include records, profiles, preferences, memories, capabilities, plans, reflections, trajectories, and policies. Tool-use ability differs from permission: revocation can block use without removing capability. \emph{Content residue} remains readable; \emph{behavioral residue} influences decisions; \emph{derivational residue} enables reconstruction through summaries, skills, or artifacts. Raw deletion alone addresses local content residue, not all three forms.

\smallskip \noindent \textbf{Demonstrated Results.} Table~\ref{tab:agent-evidence} groups studies by role. Three direct cases differ in target, state, protocol, and claim: ToolDelete tests learned tool use, SBU coordinates parameter and memory changes, and the memory audit tests extraction after deletion. Their outcomes and limitations ground the discussion of broader requirements.

\smallskip \noindent \textbf{Learned Tool Use.} ToolDelete changes parameters to remove selected tool capabilities while preserving retained tools and general tasks~\cite{icml/ChengA25}. It aligns forget-tool responses with a tool-free model and uses task arithmetic to preserve general knowledge. Experiments include ToolAlpaca 7B, trained on tool demonstrations, and evaluate tool-dependent tasks, general benchmarks, and LiRA-Tool membership inference using generated shadow queries. On ToolAlpaca, deleting 20\% of tools with the DPO variant reduces forget-tool performance from 75.7 to 28.7 while retained-tool performance changes from 73.1 to 73.3; general performance changes from 24.1 to 23.1. Sequential deletion also retains performance close to one-shot deletion. These results concern learned capability under the evaluated queries, rather than permission revocation. They do not test whether earlier tool calls left target information in files, messages, or external services, where removal would require tracing those outputs and verifying changes to the affected external records.

\smallskip \noindent \textbf{Parameters and Memory.} SBU targets medical QA information in II-Medical-8B with a 2,000-entry memory and top-five retrieval~\cite{corr/Wang2602Agentic}. It blocks and removes target memories and handles dependent artifacts before updating parameters against a random reference while retaining QA performance. Evaluation combines forget, test, and generalization accuracy, membership inference, memory-side tests, and an interaction sequence of storage, querying, deletion, and probing. Ablations compare the joint method with parameter-only and memory-only variants. On MedQA with 100 forget items, the joint method lowers membership-inference AUC from 0.637 to 0.552 while test accuracy rises from 91.0\% to 92.5\%; forget accuracy remains 73.0\%. The interaction test reports zero target retrieval hits after deletion, with retained memories still accessible. Thus retrieval exclusion and residual answering support different conclusions even within this configuration. The results support coordination in a controlled prototype, but do not establish complete target removal or sustained protection across independently owned agents and later replica reconnection.

\smallskip \noindent \textbf{Residual Memory After Deletion.} The memory-residue audit targets synthetic canaries inserted into 50 LongMemEval histories, using Gemma 3 12B and a partial replication with GPT-4o-mini~\cite{corr/Lei2606Agents}. It changes raw and summary memory, including text and embeddings, rather than model parameters. Direct, indirect, and jailbreak probes measure verbatim extraction before and after deletion; personalization recall measures utility. With key-fact summaries, raw-only deletion leaves worst-tier residue of 0.20 on Gemma and 0.22 on GPT-4o-mini. Re-summarization reduces these values to 0.11 and 0.10, while full purge and tombstone redaction yield zero in the tested settings. This comparison directly exposes surviving derivatives as a source of leakage. Its scope remains synthetic canaries, two memory tiers, and the declared probes; it does not test later interactions or updates that could restore the target. Unlike ToolDelete, the intervention removes stored records; unlike SBU, it does not coordinate memory deletion with parameter updates.

\smallskip \noindent \textbf{Related Mechanisms and Settings.} CLPU-DER++ studies task removal in lifelong agents~\cite{collas/LiuLS22}, while Reinforcement Unlearning and TrajDeleter address environments and trajectories in sequential policies~\cite{ndss/YeZZWGSSZ25,ndss/GongLYW25}. These studies inform target construction but do not evaluate an LLM-agent stack. ALU uses agents to control another model at inference time~\cite{colm/SanyalM25}; Secure Forgetting translates requests into behavioral unlearning prompts~\cite{corr/Ye2604Secure}. Editable memory architectures~\cite{acl/YuYXTFLLW26,acl/LatimerBNBSWR26} and memory-management studies~\cite{acl/XiongLXHLTLX26} supply interfaces and utility evidence, while pruning and curation address retention objectives~\cite{corr/Gu2604FSFM,corr/Wu2606ForgetImprove}. Their relevance depends on what they change and test: an available deletion interface or improved utility alone does not demonstrate removal of the requested influence.

\begin{table*}[!t]
\caption{Representative agentic studies by role in this review, tested outcome, and claim boundary. A dagger marks preprints; roles describe their use here, not mutually exclusive paper categories.}
\label{tab:agent-evidence}
\centering
\scriptsize
\setlength{\tabcolsep}{4pt}
\renewcommand{\arraystretch}{1.06}
\arrayrulecolor{stageblue!65}
\begin{tabularx}{\textwidth}{L{0.14\textwidth}L{0.27\textwidth}L{0.27\textwidth}Y}
\toprule
\rowcolor{stageblue!16}
\textbf{Evidence role} & \textbf{Representative work} & \textbf{Tested outcome / mechanism} & \textbf{Claim limitation} \\
\midrule
\cellcolor{stageblue!8}Adjacent & CLPU-DER++~\cite{collas/LiuLS22} & Task removal & Not an LLM-agent stack \\
\rowcolor{stageblue!4}
\cellcolor{stageblue!8}Adjacent & TrajDeleter~\cite{ndss/GongLYW25}; RL unlearning~\cite{ndss/YeZZWGSSZ25} & Trajectory; environment & Sequential-policy setting \\
\cellcolor{stageblue!8}Direct evidence & ToolDelete~\cite{icml/ChengA25} & Reduced learned tool use & Prior outputs untested \\
\rowcolor{stageblue!4}
\cellcolor{stageblue!8}Direct evidence & SBU$^\dagger$~\cite{corr/Wang2602Agentic} & Joint parameter-memory tests & Controlled medical QA prototype \\
\cellcolor{stageblue!8}Direct evidence & Memory-residue audit$^\dagger$~\cite{corr/Lei2606Agents} & Memory-tier leakage after deletion & No longitudinal recurrence test \\
\rowcolor{stageblue!4}
\cellcolor{stageblue!8}Supporting & ALU~\cite{colm/SanyalM25} & Inference-time response control & Mediators, not forget targets \\
\cellcolor{stageblue!8}Supporting & Secure Forgetting$^\dagger$~\cite{corr/Ye2604Secure} & Request-conditioned agent behavior & Distributed removal untested \\
\cellcolor{stageblue!8}Supporting & Memory management~\cite{acl/XiongLXHLTLX26} & Deletion changes later utility & No target-removal test \\
\rowcolor{stageblue!4}
\cellcolor{stageblue!8}Supporting & Memory systems~\cite{acl/YuYXTFLLW26,acl/LatimerBNBSWR26} & Editable persistent state & Residual influence untested \\
\cellcolor{stageblue!8}Adjacent & Pruning and curation$^\dagger$~\cite{corr/Gu2604FSFM,corr/Wu2606ForgetImprove} & Retention management & Objective differs from removal \\
\bottomrule
\end{tabularx}
\end{table*}

\smallskip \noindent \textbf{Requirements Beyond Demonstrated Results.} These studies do not jointly test plans, tool outputs, shared memory, and independent agents. The following requirements concern this untested combination and motivate Section~\ref{sec:challenges}. They identify dependencies and observations needed for broader claims across models, memory, and external replicas.

Plans and skills can depend on an observation without retaining its wording. Provenance can guide deletion or recomputation; matched replay can test resulting decisions while disclosing missing lineage. Tool effects add a separate limit: capability removal does not erase earlier files, messages, or service records. Compensation addresses records without undoing disclosure. Reports distinguish verified removal, suppression, external requests, and unverifiable state.

Messages, shared memory, and replicas can restore a target after local removal. Coordinated protocols therefore need dependency tracking, versioned changes, and tests after reconnection or updates. Acknowledgments cannot replace inspection or behavioral tests. Independently owned components may permit only notification or limited checks, rather than the access available within $B_r^{\mathrm{ctrl}}$. Their unresolved paths remain part of the requested boundary.

\smallskip \noindent \textbf{Reporting Path Coverage.} To make these proposed requirements explicit, our framework distinguishes completed tests from passed residual tests; this is a reporting rule, not an empirical result of the studies above. For a nonempty path set $\mathcal{P}_r$ with positive weights $w_p$, define:
\begin{equation}
\mathrm{Cov}_{\mathrm{test}}(r)
 =\frac{\sum_{p\in\mathcal{P}_r}w_p z_p}{\sum_{p\in\mathcal{P}_r}w_p},
\qquad
\mathrm{Cov}_{\mathrm{pass}}(r)
 =\frac{\sum_{p\in\mathcal{P}_r}w_p z_p b_p}{\sum_{p\in\mathcal{P}_r}w_p}.
\label{eq:path-coverage}
\end{equation}
Here $z_p$ is the completed-test indicator defined in Section~\ref{sec:claim-definitions}. Set $b_p=1$ if $z_p=1$ and $V_{p,a,h}\leq\epsilon_p$ for every declared $(a,h)$, and $b_p=0$ otherwise. Indicator $z_p$ distinguishes an untested path from a completed test; together, $(z_p,b_p)$ distinguish untested, failed, and passed residual evaluations. Thus $0\leq\mathrm{Cov}_{\mathrm{pass}}\leq\mathrm{Cov}_{\mathrm{test}}\leq1$ for the declared path inventory.

The interpretation depends on the weighting rule. Uniform weights report a path fraction; risk weights require a disclosed policy. Comparisons require a common path decomposition and weighting rule: subdividing routes can change the denominator without changing the system.
These ratios report coverage, not success probabilities or compositional guarantees. An empty inventory is reported as not evaluated, rather than assigned perfect coverage. Neither score validates the inventory, substitutes for specificity or utility, or proves persistence beyond tested times. Uncontrolled required paths remain limitations on the claim, even when every inspected component passes.

\begin{tcolorbox}[colback=softblue,colframe=stageblue!70,boxrule=0.8pt,arc=3pt]
\textbf{Takeaway:} Removal beyond the model depends on the state changed and paths tested. Studies demonstrate reduced tool use, coordinated parameter and memory interventions, and leakage from surviving summaries. Broader claims require tracing dependencies through retrieval, tools, and shared state, then testing propagation and restoration over time. Inaccessible paths and untested interactions constrain the resulting removal claims.
\end{tcolorbox}

%% file: figures/agent_anatomy_agentic.tex
\includegraphics[width=\textwidth]{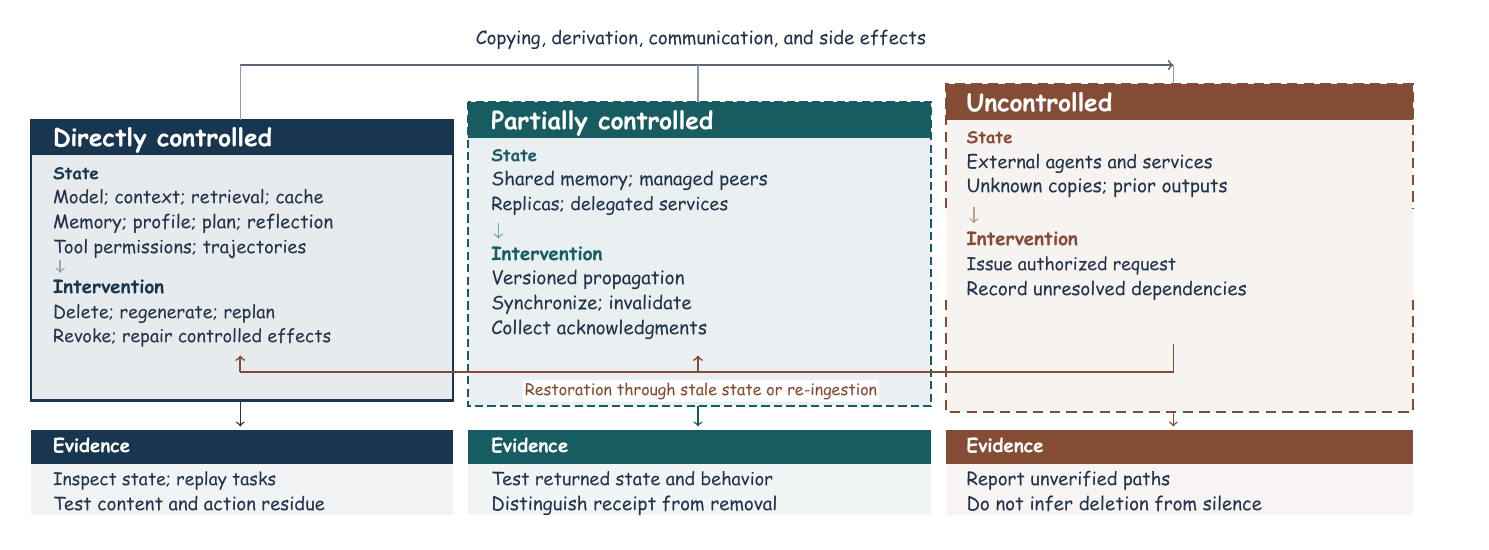}

%% file: sections/08_evaluation_verification.tex
\section{Evaluation, Verification, and Guarantees}
\label{sec:system-evaluation}

Section~\ref{sec:claim-definitions} defines the evidence profile and empirical decision rule. Here we explain how to implement that rule: choose references, combine local inspections with end-to-end tests, control recovery experiments, and schedule later observations. Table~\ref{tab:evaluation-axes} translates model-level tests into the additional checks needed for system state and behavior.
\begin{table*}[!t]
\caption{Illustrative model-level tests and system-level obligations across six evidence dimensions.}
\label{tab:evaluation-axes}
\centering
\scriptsize
\setlength{\tabcolsep}{4pt}
\renewcommand{\arraystretch}{1.06}
\arrayrulecolor{stageblue!65}
\begin{tabularx}{\textwidth}{L{0.15\textwidth}L{0.34\textwidth}Y}
\toprule
\rowcolor{stageblue!16}
\textbf{Dimension} & \textbf{Model-level evidence} & \textbf{System-level extension} \\
\midrule
\cellcolor{stageblue!8}Efficacy & Target recall; likelihood; extraction & Target disclosure or action via $V_{p,a,h}$ \\
\rowcolor{stageblue!4}
\cellcolor{stageblue!8}Specificity & Retained neighbors; related facts & Non-target state and behavior via $D_h$ \\
\cellcolor{stageblue!8}Utility & Reasoning; safety; language quality & Authorized end-to-end tasks via $Q_h$ \\
\rowcolor{stageblue!4}
\cellcolor{stageblue!8}Robustness & Prompt attacks; probes; relearning & Alternate tools and access in $\mathcal{A}_r$ \\
\cellcolor{stageblue!8}Persistence & Retests after subsequent changes & Evolution under $\tau_r$ at times $\mathcal{H}_r$ \\
\rowcolor{stageblue!4}
\cellcolor{stageblue!8}Completeness & Required loci and access paths & Closure inventory and path-test coverage \\
\bottomrule
\end{tabularx}
\end{table*}
The target, required paths, and preservation criteria determine the tests and references in Table~\ref{tab:evaluation-axes}. Component inspections establish state changes; behavioral tests assess their effects on target disclosure and retained tasks across components.

\smallskip \noindent \textbf{Baselines and Counterfactuals.} Retraining without the forget set is stochastic and often infeasible. It provides a reference for training-data removal; the original model establishes prior target presence. Gradient or preference updates, parameter-efficient methods, guardrails, and editing provide alternative intervention baselines. Their role depends on the outcome compared: refusal tests exposure control~\cite{nips/CooperEtAl25}, while editing can benchmark factual revision and residual prior knowledge~\cite{aaai/LiWSKQCWL26}. For external state, replay without the target event provides a counterfactual under the assumptions in Section~\ref{sec:foundations}. Retained-task and related-fact tests separate intended changes from collateral damage.

\smallskip \noindent \textbf{Component-Level and End-to-End Tests.} Component tests establish whether local state changed as required. End-to-end tests then examine whether the combined system still discloses, reconstructs, or uses the target. A path matrix crosses $p\in\mathcal{P}_r$ with access conditions $a\in\mathcal{A}_r$ and observation times $h\in\mathcal{H}_r$. Cells record baseline, transformation, residual score, uncertainty, and pass, fail, or untested status. Equation~\ref{eq:path-coverage} summarizes coverage without hiding untested paths. Augmented systems also test whether authorized context remains usable under the declared policy while distinguishing permitted contextual use from forbidden recall of target information retained in model parameters~\cite{icml/PengAGBHWH26}.

\smallskip \noindent \textbf{Recovery Attacks.} Recovery evidence spans adversarial prompts~\cite{iclr/PatilHB24,tmlr/LuckiWH0TR25,emnlp/ToL25}, relearning~\cite{iclr/0001FWS25}, quantization~\cite{iclr/ZhangWLWTL00W25}, reasoning~\cite{corr/sinha2025step}, repeated sampling~\cite{icml/ReisizadehRCPLH26}, and privacy attacks~\cite{icml/DuWZPHR25,nips/WuPLW25}. To avoid mistaking attack-induced outputs for recovery, audits require negative controls for strings never learned~\cite{emnlp/ChenSXH25} and controls that distinguish syntactic from topical relearning~\cite{iclr/Yoon2026RethinkingBR,acl/ChangL25}. REMIND probes black-box loss geometry~\cite{acl/CohenNM26}. The resulting recovery scores should be reported with attack access, sampling budgets, and the reference used to identify residual target knowledge.

\smallskip \noindent \textbf{State Changes and Maintenance Tests.} Relearning and quantization create separate evaluation branches from the observed state. Each branch requires an equally transformed reference, and any claim of preserved utility requires utility tests on that branch. Other attack conditions start separately, so that one transformation does not silently affect the next test. For persistence, declare later observations in $\mathcal{H}_r$ and retest the earlier removal postcondition after fine-tuning, merging, further requests, index rebuilding, cache warming, trajectory replay, or replica reconnection. A transformed branch contributes persistence evidence only when it is included as such a later observation. At each observation, apply $\tau_r$ to distinguish forbidden recurrence from reintroduction authorized by the declared request policy.

\smallskip \noindent \textbf{Evaluator Validity and Collateral Effects.} Exact match misses semantic reconstruction. Knowledge-graph evaluation exposes inference through retained relations~\cite{nips/Wei2025DoLR}, and OpenUnlearning integrates 16 evaluation metrics and meta-evaluates 12~\cite{nips/DornaMZMKLM25}. A fixed retained-task suite can also miss damage: dynamic probes reveal knowledge holes~\cite{nips/KoJFMJ25}, DUSK separates shared content~\cite{acl/JeungYHKSHYN26}, and narrow removal can induce misalignment~\cite{acl/MushtaqRKSGCZG26}. Standard tasks and harnesses aid reproducibility~\cite{iclr/HendrycksBBZMSS21,nips/HendrycksBKABTS21,acl/LinHE22,corr/eval-harness24,Evalchemy25} without replacing target-specific extraction, recovery, or longitudinal recurrence tests.
Side effects can escape curated tests: MNEME probes sparse model differences~\cite{emnlp/KassemSRF25}; honesty-focused work studies hallucinated or inconsistent refusal~\cite{acl/GuDZL26,nips/TanQLZCCG25,iclr/chen2026unlearning}. LLM judges expand semantic coverage but require calibration~\cite{nips/ZhengC00WZL0LXZ23}. Per-target distributions, tail risk, uncertainty, and calibration help expose hidden damage~\cite{icml/Wei25}.

\smallskip \noindent \textbf{Security of Requests and Updates.} Recovery tests examine whether target influence survives; interface tests examine whether the removal procedure creates another route to harm. Comparisons between checkpoints can expose removed features~\cite{sp/HuWDX24}; malicious requests and prepared data can induce excessive removal, weaken refusal behavior, or activate backdoors~\cite{ndss/HuWCZSHZM24,usenix/Song25,usenix/HuangMZ24}. Protocols therefore validate request authority, specify which versions an evaluator can access, and retest retained safety after updates. Feature or label revocation and resource-constrained exact unlearning further show why target type and available operations must be stated alongside the claimed guarantee~\cite{ndss/WarneckePWR23,sp/Xia25}.

\smallskip \noindent \textbf{Evidence and Guarantees.} Behavioral, representational, adversarial, counterfactual, certified, and system-audited evidence answer different questions. A proof of distributional equivalence concerns the algorithm and domain covered by its assumptions~\cite{usenix/thudi2022necessity,nips/GinartGVZ19}; a propagation audit checks whether required copies and descendants were handled. A system combining both forms of evidence should identify how the proof's model assumptions relate to the audited deployment, including any state excluded from either assessment. This links formal and operational evidence.

\smallskip \noindent \textbf{Reporting Requirements.} Reports specify the request, boundary, versions, access, intervention, references, and prior target presence. They predeclare thresholds, budgets, and observation times, then record decoding, attacks, propagation, uncertainty, and artifacts. Compute, storage, runtime, and audit costs are reported separately from test outcomes and proof assumptions. Test records distinguish a violated criterion from an untested path or partial coverage. An empty evidence-matrix code means no qualifying test was identified in the report, leaving unreported tests unknown.

\begin{tcolorbox}[colback=softblue,colframe=stageblue!70,boxrule=0.8pt,arc=3pt]
\textbf{Takeaway:} A useful evaluation connects the request to suitable references, local inspections, and end-to-end tests. Controlled recovery experiments probe residual influence, while scheduled retests track removal through later changes. Reporting path coverage, retained performance, uncertainty, and costs makes clear which requirements were met, which failed, and which remain untested within the declared boundary.
\end{tcolorbox}

%% file: sections/09_applications_deployment.tex
\section{Application Objectives and Deployment Implications}
\label{sec:applications}

The preceding chapters compare how removal is attempted and tested. Here we distinguish two purposes for using these interventions: satisfying a target-removal requirement and improving learning or understanding model behavior. The first requires evidence that the requested influence has been removed within $B_r$; the second requires evidence that the intervention caused the claimed benefit. Table~\ref{tab:deployment-cases} connects these purposes to targets, checks, and failure modes. Access and cost constrain both; performance gains alone establish neither removal nor compliance.

\begin{table*}[!ht]
\caption{Illustrative removal targets, priority checks, and failure modes by deployment objective.}
\label{tab:deployment-cases}
\centering
\scriptsize
\setlength{\tabcolsep}{4pt}
\renewcommand{\arraystretch}{1.06}
\arrayrulecolor{stageblue!65}
\begin{tabularx}{\textwidth}{L{0.15\textwidth}L{0.21\textwidth}YY}
\toprule
\rowcolor{stageblue!16}
\textbf{Objective} & \textbf{Typical target} & \textbf{Priority checks} & \textbf{Characteristic failure} \\
\midrule
\cellcolor{stageblue!8}Privacy / deletion & Identity; profile; chats & Efficacy; completeness; robustness & Recoverable personal data \\
\rowcolor{stageblue!4}
\cellcolor{stageblue!8}Copyright & Passage; document; domain & Efficacy; specificity; utility & Target expression recoverable \\
\cellcolor{stageblue!8}Safety / secrecy & Secret; behavior; capability & Robustness; efficacy; utility & Suppressed but recoverable skill \\
\rowcolor{stageblue!4}
\cellcolor{stageblue!8}Corrective removal & Bias; label errors & Specificity; utility; efficacy & Harm shifted to other groups \\
\cellcolor{stageblue!8}Knowledge update & Outdated fact or memory & Persistence; efficacy; specificity & Revoked value remains actionable \\
\rowcolor{stageblue!4}
\cellcolor{stageblue!8}Agent governance & Memory; tool; shared state & Completeness; persistence; utility & Peer or replica restores target \\
\cellcolor{stageblue!8}Functional study & Step; response; domain & Utility; specificity; matched controls & Benefit misattributed to removal \\
\bottomrule
\end{tabularx}
\end{table*}

\subsection{Applications Aimed at Target Removal}

\smallskip \noindent \textbf{Privacy, Copyright, and Safety.} For privacy requests concerning people, provenance and authority matter alongside behavioral change; regulatory language does not define a technical postcondition~\cite{mantelero2013,harding2019}. Tests should cover aliases, indirect attributes, semantic neighbors, and minority risk~\cite{icml/Wei25}. Copyright targets span passages, entities, and domains. WHP and MUSE cover literary sources~\cite{corr/eldan23,iclr/shi24}, while related work distinguishes output evidence from training provenance and warns that sparse forget sets can omit related passages, entities, and other in-scope content~\cite{emnlp/KaramolegkouLZS23,coling/MaFZH0FQ25}.

For harmful or secret knowledge, ordinary accuracy must be supplemented by adversarial recovery tests. WMDP and RMU operationalize hazardous-domain forgetting~\cite{icml/LiPGYBGLDGMHLJL24}; backdoor and jailbreak defenses supply narrower corrective cases, while the same intervention can create ripple effects and collateral vulnerabilities~\cite{aaai/JiangLLM25,corr/zhang2024theft,icml/youssef2502}. Recovery and retained-safety tests assess removal, residual exposure, and damage to retained protections~\cite{corr/Zou23,acl/JiHZ0DZQZWLHG025}.

\smallskip \noindent \textbf{Fairness, Bias, and Corrective Unlearning.} Fairness can motivate or constrain removal. Bias attenuation targets predictive shortcuts~\cite{acl-deeplo/HeZW19}; fairness-aware deletion limits disparities during removal~\cite{aistats/OesterlingMCL24,corr/Ravindra23}. Continual bias correction and noisy-label unlearning--relearning provide adjacent evidence~\cite{kdd/0001G0WLW025,ijcai/Sui00NL025}, not demonstrated benefits across LLM deployments. Requests specify counterfactuals, retained-utility thresholds, and acceptable group-level collateral change.

\smallskip \noindent \textbf{Knowledge Updates and Temporal Consistency.} Updates reconcile conflicting values. Facts can become stale or differ across model versions~\cite{emnlp/PetroniRRLBWM19,corr/Cheng24}; continual updates can interfere with prior state~\cite{iclr/JangYYSHKCS22}, and RAG remains vulnerable to temporal inconsistency~\cite{naacl/LiLRZNA25,acl/LiuCL26}. Externalizing facts improves editability~\cite{iclr/ZhaoZBLZNGWAS26}, but that design choice does not determine whether old values may remain in historical records. Policy $\tau_r$ must distinguish erasure, deprioritization, versioning, and authorized retention; otherwise, a valid historical response may be mistaken for target removal failure.

\smallskip \noindent \textbf{Applying These Objectives to Multimodal Systems.} Studies test single-image removal from multimodal LLMs~\cite{nips/LiWZQD0BL24}, continual vision and generation forgetting~\cite{CVPR/zhao2024,icmlw/lee2025an}, and localized revision~\cite{cvpr/GolatkarAS20}. Image-generation work emphasizes preservation, robustness, and auditing~\cite{ccs/Liu25,ccs/Liu252}. These findings inform evaluation of multimodal components within $B_r$ under the studied access conditions and protocols, but do not establish removal in text-based agents.

\input{sections/09a_functional_forgetting}

\begin{tcolorbox}[colback=softblue,colframe=stageblue!70,boxrule=0.8pt,arc=3pt]
\textbf{Takeaway:} Application objectives determine authorized targets, retention policies, acceptable risk, and evidence requirements. Privacy, safety, fairness, and knowledge updates require different checks. Functional gains require matched controls and do not establish target removal. Intervention choice follows system access and dependencies; technical scores alone establish neither authorization nor compliance in deployed systems.
\end{tcolorbox}

%% file: sections/09a_functional_forgetting.tex
\subsection{Applications Aimed at Learning and Analysis}\label{sec:functional}

Functional forgetting uses selective weakening or removal to probe model behavior, support learning, or investigate regularization. Here the intended outcome is an analytical insight or performance benefit, which changes the final layer of the comparison: evidence must attribute that outcome to the intervention using suitable controls. A removal claim requires separate tests of whether the designated target's influence remains recoverable under the declared conditions.


\smallskip \noindent \textbf{Mechanism Probe.} Chain-of-thought (CoT) can improve multi-step performance~\cite{nips/wei2022,colm/Zeng2025,icml/edward2025}, but an emitted rationale need not identify the computation responsible for an answer. Selective intervention on a reasoning step can test how the answer distribution changes~\cite{emnlp/tutek2025}. A shift supports sensitivity to the intervention; stability can indicate redundancy, compensation, or an irrelevant step. Causal interpretation requires localized changes and matched controls that rule out general knowledge loss, unintended optimization damage, and compensating mechanisms~\cite{corr/xie24}.

\smallskip \noindent \textbf{Learning Aid.} Biological forgetting offers conceptual background~\cite{anderson2021active}. In graceful forgetting, Jiang et al.~\cite{emnlp/jiang25} rank responses for periodic unlearning alongside positive updates. Continual vision and noisy-label correction offer adjacent evidence~\cite{kdd/0001G0WLW025,ijcai/Sui00NL025}, not direct LLM benefits. To attribute a benefit to the forgetting intervention, comparisons must match budgets and data exposure between the intervention and an appropriate learning baseline and assess adaptation, retained tasks, and response attenuation separately. Unintended forgetting does not prove targeted removal~\cite{iclr/ZhengCQ025}.

\smallskip \noindent \textbf{Regularization Perspective.} Performance gains under random or incorrect mathematical rewards~\cite{corr/shao2025,corr/wu2025} raise questions about whether improvements reflect correct mathematical learning or other optimization effects. These studies motivate testing selective weakening as a regularizer, without testing unlearning. Matched learning baselines, equal data exposure, and separate attenuation and utility tests are needed to attribute gains to regularization.

%% file: sections/10_trends_evidence.tex
\section{Evidence Synthesis and Comparative Findings}
\label{sec:trends}
We compare recurring findings and conflicting results across the preceding chapters, separating target, state, and protocol differences before interpreting rankings. Section~\ref{sec:challenges} then examines unresolved problems and their evaluation.

\subsection{Review Methodology}
\label{sec:review-methodology}

This structured narrative review covers work through 25 August 2026, drawing on mapping and citation snowballing guidance~\cite{ease/PetersenFMM08,ease/Wohlin14}. Review notes list ACM Digital Library, IEEE Xplore, Scopus, Web of Science, DBLP, ACL Anthology, PMLR, OpenReview, arXiv, and conference proceedings and programs as sources. Query families combine unlearning, forgetting, data deletion, knowledge removal, and the right to be forgotten with LLMs, retrieval, context, embeddings, caches, memory, tools, trajectories, and agents. Complete queries, dated search exports, and screening counts are unavailable, so source-specific searches cannot be reconstructed and exhaustive coverage is not claimed.

We include studies that test target removal or inform its mechanisms, evaluation, and boundaries. Ordinary forgetting and retention management are included only when relevant to this scope. Direct evidence tests removal; supporting work provides mechanisms; adjacent research informs definitions, threats, or system constraints. These roles may overlap. Preprint and published versions count as one contribution, with claims tied to the cited version.

Table~\ref{tab:survey-comparison} codes explicit review coverage, crediting empirical synthesis and analytical discussion under the same rules for every survey. Table~\ref{tab:evidence-matrix} instead codes reported tests in particular settings against the definitions in Section~\ref{sec:claim-definitions}. We assign dedicated, partial or proxy, and unidentified-test codes from the described protocols, then check them against the cited reports and compare like cases for consistency. The discussion below explains the main judgments. This is a qualitative coding procedure; independent screening agreement and experimental reproduction are not claimed.

\subsection{Comparison of Reported Evidence}
\label{sec:evidence-comparison}

\begin{table*}[!t]
\caption{Evidence profiles of selected studies.}
\label{tab:evidence-matrix}
\centering
\footnotesize
\setlength{\tabcolsep}{2pt}
\renewcommand{\arraystretch}{1.16}
\arrayrulecolor{stageblue!65}
\begin{tabularx}{\textwidth}{@{}L{.25\textwidth}*{5}{>{\centering\arraybackslash}X}C{.14\textwidth}@{}}
\toprule
\rowcolor{stageblue!16}
\tablecolorleft{stageblue!16}\textbf{Study / setting} & \textbf{Efficacy} & \textbf{Specificity} & \textbf{Utility} & \textbf{Robustness} & \textbf{Persistence} &\tablecolorright{stageblue!16} \textbf{Completeness} \\
\midrule
\rowcolor{stageblue!9}[0pt][0pt]
\multicolumn{7}{@{}c}{\textit{Standalone LLMs: benchmark and recovery evidence}} \\
\tablecolorleft{stageblue!8}TOFU~\cite{colm/Maini24} & \fullcirc & \halfcirc & \fullcirc & \emptycirc & \emptycirc & \emptycirc \\
\rowcolor{stageblue!4}
\tablecolorleft{stageblue!8}WMDP / RMU~\cite{icml/LiPGYBGLDGMHLJL24} & \fullcirc & \halfcirc & \fullcirc & \fullcirc & \halfcirc &\tablecolorright{stageblue!4} \emptycirc \\
\tablecolorleft{stageblue!8}MUSE~\cite{iclr/shi24} & \fullcirc & \halfcirc & \fullcirc & \halfcirc & \emptycirc & \emptycirc \\
\rowcolor{stageblue!4}
\tablecolorleft{stageblue!8}OpenUnlearning~\cite{nips/DornaMZMKLM25} & \fullcirc & \halfcirc & \fullcirc & \fullcirc & \halfcirc &\tablecolorright{stageblue!4} \emptycirc \\
\tablecolorleft{stageblue!8}Leak@$k$~\cite{icml/ReisizadehRCPLH26} & \fullcirc & \emptycirc & \fullcirc & \fullcirc & \emptycirc & \emptycirc \\
\midrule
\rowcolor{stageblue!9}[0pt][0pt]
\multicolumn{7}{@{}c}{\textit{Extended loci: retrieval-conditioned behavior, learned tools, and memory}} \\
\tablecolorleft{stageblue!8}RAG control$^\dagger$~\cite{corr/Wang2410RAG} & \fullcirc & \halfcirc & \fullcirc & \fullcirc & \emptycirc & \emptycirc \\
\rowcolor{stageblue!4}
\tablecolorleft{stageblue!8}ToolDelete~\cite{icml/ChengA25} & \fullcirc & \halfcirc & \fullcirc & \halfcirc & \halfcirc &\tablecolorright{stageblue!4} \emptycirc \\
\tablecolorleft{stageblue!8}SBU$^\dagger$~\cite{corr/Wang2602Agentic} & \fullcirc & \halfcirc & \fullcirc & \halfcirc & \emptycirc & \halfcirc \\
\rowcolor{stageblue!4}
\tablecolorleft{stageblue!8}Memory residue$^\dagger$~\cite{corr/Lei2606Agents} & \fullcirc & \emptycirc & \halfcirc & \fullcirc & \emptycirc & \halfcirc \\
\bottomrule
\end{tabularx}
\par\smallskip
\begin{minipage}{\textwidth}
\scriptsize
\fullcirc\ Dedicated tests within the evaluated setting;
\halfcirc\ Partial tests or a limited proxy;
\emptycirc\ No qualifying test identified.
$\dagger$ Preprint.
\end{minipage}
\end{table*}

Table~\ref{tab:evidence-matrix} reveals uneven coverage across the selected studies: efficacy and utility are widely tested in these settings, whereas later observations and path inventories are less developed. Its two blocks compare checkpoint and extended-locus evaluations, and Figure~\ref{fig:literature-map} places them within the wider literature. The entries describe reported test coverage, including tests that expose failures, rather than performance rankings or an exhaustive census of work in the field.

The codes distinguish the question tested from the result obtained. A direct extraction or recovery experiment counts as a dedicated robustness test even when it succeeds in recovering the target; membership inference alone provides a narrower proxy. Thus ToolDelete's LiRA-Tool supports partial robustness coverage, whereas the memory audit directly tests canary extraction. The audit's raw and summary tiers provide partial completeness evidence, but its utility measurements across memory configurations do not fully establish preservation after deletion. These judgments explain why dedicated tests in one dimension can coexist with partial or unidentified tests in another.

Persistence coding asks whether an earlier target is retested after state changes. WMDP/RMU evaluates WMDP-Cyber after fine-tuning on the forget corpus; OpenUnlearning compares target metrics before and after relearning and 4-bit quantization. These transformation tests provide limited proxies for persistence, not observation over a deployment horizon. Re-exposure to forget data is interpreted as an attack condition, not automatically a policy violation. ToolDelete reports aggregate forget-tool performance after sequential batches; this is partial coverage because earlier batches are not tracked separately. The codes describe these tests, not successful or durable removal.

MUSE tracks retained utility across requests, not earlier target removal, so its persistence cell remains empty. Leak@$k$ samples a fixed checkpoint, and the memory audit tests deletion across memory tiers without later updates. SBU adds parameter and memory tests, retained QA, membership inference, and summary-cleanup checks~\cite{corr/Wang2602Agentic}. These support dedicated efficacy and utility tests, with partial specificity, robustness, and completeness coverage. Its Store--Query--Delete--Probe sequence checks the deletion outcome, but does not establish retesting after a subsequent state-changing operation; no qualifying persistence test is coded. All three direct agentic cases are coded consistently.

Preservation tests show a related mismatch in resolution. Broad retained-task scores are common, but they need not isolate damage to facts close to the target, explaining the partial specificity codes. Likewise, no qualifying inventory and coverage audit was identified for the checkpoint evaluations, including ToolDelete; its learned-tool tests address a different boundary from earlier external tool outputs. ERASE's knowledge-base consistency tests~\cite{naacl/LiLRZNA25} and utility evaluations after memory deletion~\cite{acl/XiongLXHLTLX26} complement this evidence by examining how external state remains usable. Their consistency and utility results do not establish that removal persists after later updates and subsequent interactions.
\input{figures/literature_tree}

\subsection{Patterns Across Studies}

\smallskip \noindent \textbf{Target Semantics.} Targets now span examples, templated biographies, entities, copyrighted domains, hazardous capabilities, reasoning traces, and overlapping knowledge~\cite{nips/JinCWHYL00024,iclr/shi24,coling/MaFZH0FQ25,emnlp/Wang2025ReasoningMU,acl/JeungYHKSHYN26}. Removability depends on salience, training origin, and granularity~\cite{acl/BorisiukSPT26,emnlp/WanRCGCG25,acl/YoonKJ26}. Together, these studies indicate that differences in target construction can explain changes in both removal and collateral damage, even under the same intervention and evaluation protocol.

\smallskip \noindent \textbf{Selective Updates and Their Requirements.} Constrained objectives and gradient correction manage forget--retain conflict~\cite{nips/Fan25,nips/EntesariHKRF25,icml/WangWFLHDH25}, while low-rank, token, and feature selection restrict where updates act~\cite{acl/XiaoMCYZYCC26,acl/LeeLX26,icml/LuPSNTL26,emnlp/WenFGWLSGS25,emnlp/WangWLHZ25}. These choices explain different results: NPO varies the loss balance, ReGLU changes the update subspace under the same loss, and PISCES relies on selected SAE features. Thus selectivity depends on data and preparation as well as the update rule. Section~\ref{sec:standalone-methods} compares how these requirements affect reported removal, preservation, and recovery results.

\smallskip \noindent \textbf{Beyond Aggregate Scores.} Semantic, capability, and shared-knowledge evaluations expose differences hidden by average accuracy~\cite{emnlp/ChengHW25,acl/LangGHZZYSZ25,acl/JeungYHKSHYN26}. Recovery through prompts, sampling, quantization, reasoning, or checkpoint comparisons can further reverse apparent success~\cite{emnlp/ToL25,icml/ReisizadehRCPLH26,icml/DuWZPHR25,nips/WuPLW25,iclr/ZhangWLWTL00W25,corr/sinha2025step}. However, probes can also introduce false positives~\cite{emnlp/ChenSXH25}. Thus conflicting evaluations can reflect either missed residual influence or artifacts of the test, requiring suitable negative controls and reference comparisons before attributing the observed outputs to surviving target influence.

\smallskip \noindent \textbf{Risks Introduced by Removal.} Checkpoint comparisons, prepared data, and malicious requests show that removal can expose information or damage safety behavior~\cite{sp/HuWDX24,usenix/HuangMZ24,usenix/Song25}. These findings differ from recovery of surviving knowledge: they concern harm created by the update procedure. As Section~\ref{sec:system-evaluation} explains, interpreting update quality therefore requires both removal tests and checks of retained protections under the stated version-access conditions.

\smallskip \noindent \textbf{Repeated Requests and Extended State.} Sequential updates, adapters, and gates address request streams~\cite{naacl/DouLLDW25,icml/wuerkaixi2025adaptive,iclr/gao25,acl/0019DTLZLGW0H25,corr/xu2026pch}, but per-update efficiency does not establish affordable cumulative storage or verification. Evidence also becomes less integrated beyond model checkpoints: augmented studies test retrieval or adaptation state, while agent studies examine tools, parameter--memory updates, and memory residue~\cite{corr/Wang2410RAG,icml/MuresanuTZP25,icml/ChengA25,corr/Wang2602Agentic,corr/Lei2606Agents}. These distinct targets explain the fragmented evidence descriptions in Figure~\ref{fig:evolution}; they do not establish sustained system-wide removal.

\smallskip \noindent \textbf{What Explains Differences Across Studies.} Three findings recur: target construction affects measured removal, broad utility scores can miss local damage, and recovery tests can overturn apparent success. Differences in target origin, model knowledge, access, and protocols therefore limit method rankings. For extended systems, independently tested components do not establish coordinated removal over time. Section~\ref{sec:challenges} examines these remaining problems.

%% file: figures/literature_tree.tex
\begingroup
\newcommand{\litrow}[4]{%
  \node[leaf,draw=#4!50,fill=#4!6] (#1) at (.275,0 |- nextrow) {#3};
  \node[branch,draw=#4!50,fill=#4!9] (#1label) at (.114,0 |- #1.center) {#2};
  \draw[treeline,line width=.45pt] (#1label.east)--(#1.west);
  \coordinate (nextrow) at ([yshift=-.8pt]#1.south west);
}
\newcommand{\litgroup}[5]{%
  \path (0,0 |- #2.north)--(0,0 |- #3.south)
    node[midway,scope,fill=#4!11,draw=#4!65] (#1) {#5};
  \draw[#4!65,line width=.6pt] (#1.east)--(.105,0 |- #1.east);
  \draw[#4!65,line width=.6pt] (.105,0 |- #2.center)--(.105,0 |- #3.center);
}
\tikzset{
  leaf/.style={anchor=north west,font=\fontsize{7.5}{8}\selectfont,
    text=treeink,align=left,text width=.707\textwidth,
    inner xsep=3pt,inner ysep=.7pt,minimum height=14pt,rounded corners=2pt,line width=.4pt},
  branch/.style={anchor=west,font=\sffamily\bfseries\fontsize{6.6}{6.8}\selectfont,
    text=treeink,align=left,text width=.133\textwidth,
    inner xsep=3pt,inner ysep=0pt,minimum height=14pt,rounded corners=2pt,line width=.4pt},
  scope/.style={anchor=west,font=\sffamily\bfseries\fontsize{6.5}{7.5}\selectfont,
    text=treeink,align=center,text width=.092\textwidth,
    inner xsep=2pt,inner ysep=4pt,rounded corners=2pt,line width=.5pt}
}
\begin{figure*}[p]
\centering
\captionsetup{skip=3pt}
{\fontsize{8}{9}\selectfont\textbf{(a) Standalone LLMs: benchmarks and intervention families}\par}\smallskip
\begin{tikzpicture}[x=\textwidth,y=1pt]
\coordinate (nextrow) at (0,0);
\litrow{b1}{Predefined\\benchmarks}{MUSE~\cite{iclr/shi24}; TOFU~\cite{colm/Maini24}; WMDP~\cite{icml/LiPGYBGLDGMHLJL24}; RWKU~\cite{nips/JinCWHYL00024}; WHP$^\dagger$~\cite{corr/eldan23}; ArXiv/GitHub/Books~\cite{acl/YaoCDNWCY24}; KnowUnDo~\cite{emnlp/TianL0LWS0C024}; WPU~\cite{emnlp/LiuZJC24}; LUME~\cite{emnlp/RamakrishnaWJCBVCHG25LUME}; SemEval Task 4~\cite{semeval/RamakrishnaWJCBVCHG25}; DUAL~\cite{acl/BorisiukSPT26}; DUSK~\cite{acl/JeungYHKSHYN26}; PCH$^\dagger$~\cite{corr/xu2026pch}; ORT~\cite{corr/Ye2506Form}.}{stageblue}
\litrow{b2}{Generated\\benchmarks}{Textbook~\cite{colm/Zhu25}; BiForget~\cite{acl/xu26}.}{stageblue}
\litrow{m1}{Input / output\\intervention}{ECO~\cite{nips/Liu24}; SPUL~\cite{naacl/BhailaVW25}; Exact ICL unlearning~\cite{icml/MuresanuTZP25}; CAP~\cite{acl/WangGPPYCOLLT26}; Contextual knowledge unlearning~\cite{acl/TakashiroKGCIM25}; ICUL~\cite{icml/PawelczykNL24}; $\delta$-Unlearning~\cite{corr/Huang2404Delta}; Conformal prediction$^\dagger$~\cite{corr/Chowdhury2026}; DRAGON~\cite{icmlw/WangLLPWB25}.}{stagepurple}
\litrow{m2}{Editing and\\localization}{Task Vectors$^*$~\cite{iclr/IlharcoRWSHF23}; PISCES~\cite{emnlp/gur2025}; DSG~\cite{corr/Muhamed2504}; Editing as Unlearning~\cite{aaai/LiWSKQCWL26}; AlphaEdit$^*$~\cite{iclr/FangJWMSW0C25}; CLUE~\cite{iclr/Hang2026clue}; Mechanistic removal~\cite{icml/GuoSSED25}; KnowledgeSmith~\cite{iclr/luo2026knowledgesmith}; Span-level removal~\cite{acl/YoonKJ26}.}{stagepurple}
\litrow{m3}{Selective\\granularity}{SU~\cite{emnlp/WanRCGCG25}; Span-level removal~\cite{acl/YoonKJ26}; DTO~\cite{acl/LeeLX26}; DOGE~\cite{emnlp/WenFGWLSGS25}; MemFlex~\cite{emnlp/TianL0LWS0C024}; SatImp~\cite{icml/YangWHLZH25}; HyperUnlearn~\cite{emnlp/WangZLYL25Hyper}.}{stagepurple}
\litrow{m4}{Optimization /\\fine-tuning}{OBLIVIATE~\cite{emnlp/xu25}; RMU~\cite{icml/LiPGYBGLDGMHLJL24}; LLM Unlearning~\cite{nips/YaoXL24}; ReLearn~\cite{acl/Xu0YZDWHOC025}; RKLU~\cite{naacl/WangZSSZQ25}; POP~\cite{acl/LeeRCC24}; ULD~\cite{nips/JiLZLK0C24}; MEOW$^\dagger$~\cite{corr/GU24}; Ext-Sub~\cite{aaai/HuLHZLZ24}; GFOES~\cite{aaai/SongYXLSX26}; TIF~\cite{aaai/ZhouQZZKZ26}; R$^{2}$MU~\cite{emnlp/Wang2025ReasoningMU}; ELM~\cite{nips/Gandikota25}; WAGLE~\cite{nips/JiaLZRB024}; ERU~\cite{nips/sui2025elastic}; FLAT~\cite{iclr/WangWLPL0B0W25}; LLMEraser~\cite{iclr/DingWYL0S0025}; LoKU~\cite{iclr/ChaCHL25}; ReGLU~\cite{acl/XiaoMCYZYCC26}; Geometric removal~\cite{icml/TanLCQCG26}; RULE~\cite{nips/Zhang2025}; PRISM~\cite{iclr/Han2026dualspace}; FALW~\cite{iclr/yu2026falw}; NPO~\cite{colm/ZhangLBM24}; SimNPO~\cite{nips/Fan25}; BS~\cite{iclr/li2026llm}; ULMR~\cite{emnlp/ShiTQQNCWYQ24}; CEU~\cite{nips/EntesariHKRF25}; GRU~\cite{icml/WangWFLHDH25}; DareU~\cite{icml/LuPSNTL26}; D$^2$~\cite{icml/YangYWTHC26}; PERMU~\cite{acl/WangJHSWJLT26}; Reveal-and-Release~\cite{emnlp/XieTKWW25}; UIPE~\cite{emnlp/WangZYRRC25}; SAGO~\cite{acl/XiaoLWWYCC26}; Stable Forgetting$^\dagger$~\cite{corr/Garg2509Stable}; LLM Surgery$^\dagger$~\cite{corr/Veldanda2409Surgery}; MOUCHI$^\dagger$~\cite{corr/Akbar2409MOUCHI}; LRM forgetting$^\dagger$~\cite{corr/Le2604Reasoning}.}{stagepurple}
\litrow{m5}{Recovery-aware\\updates}{LAU~\cite{aaai/YuanJCCLZ25}; PRISM~\cite{iclr/Han2026dualspace}; SSIUU~\cite{iclr/yang2026erase}; AGTAO~\cite{acl/LiZGWDLWL26}; SSPU~\cite{emnlp/WangWLHZ25}; Attention Shifting~\cite{nips/TanQLZCCG25}; ERU~\cite{nips/sui2025elastic}; Invariance Regularization~\cite{icml/WangZJRWYPBL25}.}{stagepurple}
\litrow{c1}{Sequential\\updates}{SSU~\cite{naacl/DouLLDW25}; ALKN~\cite{icml/wuerkaixi2025adaptive}; FIT$^\dagger$~\cite{corr/xu2026pch}; FUNU$^\dagger$~\cite{corr/li2025}.}{stagecyan}
\litrow{c2}{Adapters\\and gates}{$O^{3}$~\cite{iclr/gao25}; GRUN~\cite{acl/0019DTLZLGW0H25}.}{stagecyan}
\litgroup{bench}{b1}{b2}{stageblue}{Benchmarks}
\litgroup{methods}{m1}{m5}{stagepurple}{Intervention\\families}
\litgroup{continual}{c1}{c2}{stagecyan}{Continual\\requests}
\draw[treeline,line width=.45pt] (.105,0 |- b1.center)--(b1label.west);
\draw[treeline,line width=.45pt] (.105,0 |- b2.center)--(b2label.west);
\draw[treeline,line width=.45pt] (.105,0 |- m1.center)--(m1label.west);
\draw[treeline,line width=.45pt] (.105,0 |- m2.center)--(m2label.west);
\draw[treeline,line width=.45pt] (.105,0 |- m3.center)--(m3label.west);
\draw[treeline,line width=.45pt] (.105,0 |- m4.center)--(m4label.west);
\draw[treeline,line width=.45pt] (.105,0 |- m5.center)--(m5label.west);
\draw[treeline,line width=.45pt] (.105,0 |- c1.center)--(c1label.west);
\draw[treeline,line width=.45pt] (.105,0 |- c2.center)--(c2label.west);
\end{tikzpicture}
\par\smallskip
{\fontsize{8}{9}\selectfont\textbf{(b) Evaluation and extensions beyond model state}\par}\smallskip
\begin{tikzpicture}[x=\textwidth,y=1pt]
\coordinate (nextrow) at (0,0);
\litrow{e1}{Forgetting, utility,\\and specificity}{OpenUnlearning~\cite{nips/DornaMZMKLM25}; TOFU metrics~\cite{colm/Maini24}; FPE~\cite{emnlp/ChengHW25}; UNCD~\cite{acl/LangGHZZYSZ25}; DUAL~\cite{acl/BorisiukSPT26}; DUSK~\cite{acl/JeungYHKSHYN26}; Retain analysis~\cite{acl/ChangL25}; MNEME~\cite{emnlp/KassemSRF25}; Task tests~\cite{emnlp/xu25,acl/YaoCDNWCY24,iclr/shi24}; Representation probes~\cite{icml/xu2025}; REMIND~\cite{acl/CohenNM26}; MU-Coreset~\cite{corr/Pal2504Coreset}; Full-stack$^\dagger$~\cite{corr/Fan2510Microscope}; PreUnlearn$^\dagger$~\cite{corr/Su2606PreUnlearn}.}{stageblue}
\litrow{e2}{Recovery, privacy,\\and auditing}{Relearning attacks~\cite{acl/LoBC24,iclr/0001FWS25,icml/xu2025}; LURK~\cite{emnlp/ToL25}; Prompt attacks~\cite{tmlr/LuckiWH0TR25,iclr/PatilHB24}; Leak@$k$~\cite{icml/ReisizadehRCPLH26}; RULI~\cite{usenix/Naderloui25}; Unlearning inversion$^*$~\cite{sp/HuWDX24}; Lifecycle audit~\cite{usenix/long25}; UBA-Inf$^*$~\cite{usenix/HuangMZ24}; Safety-removal attack~\cite{usenix/Song25}; Privacy Benchmark~\cite{aaai/QianZLH26}; Quantization~\cite{iclr/ZhangWLWTL00W25}; Reasoning recovery$^\dagger$~\cite{corr/sinha2025step}; Knowledge correlation~\cite{nips/Wei2025DoLR}; U-LiRA+ / TULA~\cite{icml/DuWZPHR25}; Checkpoint-pair extraction~\cite{nips/WuPLW25}; Soft-token audit~\cite{emnlp/ChenSXH25}; Trace detection~\cite{iclr/chen2026unlearning}; Minority risk~\cite{icml/Wei25}.}{stageblue}
\litrow{e3}{Operational cost\\(modifier)}{Compute~\cite{acl/YaoCDNWCY24,corr/xu2026pch,ijcnlp/KumarGR23}; Runtime~\cite{emnlp/xu25,acl/JangYYCLLS23,emnlp/ChenY23}, \cite{emnlp/WuLXDW0X23,nips/YaoXL24}; Benchmark construction~\cite{acl/xu26}.}{stageblue}
\litrow{r1}{Retrieval-time\\control / support}{RAG unlearning$^\dagger$~\cite{corr/Wang2410RAG}; ERASE$^*$~\cite{naacl/LiLRZNA25}; Personalized RAG$^*$~\cite{sigir/ShiX0Z00025}; ChunkRAG$^*$~\cite{nnacl/LewisPPPKGKLYR020}; Foundational RAG$^*$~\cite{nips/LewisPPPKGKLYR020}.}{stagecyan}
\litrow{r2}{Externalized facts}{LMLM$^*$~\cite{iclr/ZhaoZBLZNGWAS26}.}{stagecyan}
\litrow{r3}{Corpus editing}{ERASE$^*$~\cite{naacl/LiLRZNA25}; Temporal drift$^*$~\cite{acl/LiuCL26}.}{stagecyan}
\litrow{r4}{In-context\\adaptation}{Exact ICL unlearning~\cite{icml/MuresanuTZP25}; ICUL~\cite{icml/PawelczykNL24}; Contextual knowledge unlearning~\cite{acl/TakashiroKGCIM25}.}{stagecyan}
\litrow{r5}{Contextual utility}{Forget to Know, Remember to Use~\cite{icml/PengAGBHWH26}.}{stagecyan}
\litrow{a1}{Parameters\\and memory}{SBU$^\dagger$~\cite{corr/Wang2602Agentic}: joint parameter--memory unlearning.}{stagepurple}
\litrow{a2}{Behavioral and\\boundary studies}{Secure Forgetting$^{*\dagger}$~\cite{corr/Ye2604Secure} (agent behavior); CLPU-DER++$^*$~\cite{collas/LiuLS22}; Reinforcement Unlearning$^*$~\cite{ndss/YeZZWGSSZ25}; TrajDeleter$^*$~\cite{ndss/GongLYW25}.}{stagepurple}
\litrow{a3}{Memory deletion\\and residue}{Deployment memorization$^\dagger$~\cite{corr/Lei2606Agents} (residue); Periodic / history / combined deletion$^*$~\cite{acl/XiongLXHLTLX26}.}{stagepurple}
\litrow{a4}{Editable memory\\architectures}{Agentic Memory$^*$~\cite{acl/YuYXTFLLW26}; Memory-R1$^*$~\cite{acl/YanYHNDBMKKPSTM26}; Hindsight$^*$~\cite{acl/LatimerBNBSWR26}; FSFM$^{*\dagger}$~\cite{corr/Gu2604FSFM}; Budget-curated memory$^{*\dagger}$~\cite{corr/Wu2606ForgetImprove}.}{stagepurple}
\litrow{a5}{Tool capabilities}{ToolDelete; LiRA-Tool membership audit~\cite{icml/ChengA25}.}{stagepurple}
\litrow{d1}{Agent-mediated\\orchestration}{ALU$^*$~\cite{colm/SanyalM25}: inference control, not distributed removal.}{teal}
\litrow{d2}{Distributed-state\\removal}{Joint removal across messages, shared memory, peers, replicas, and services remains open; orchestration does not establish peer forgetting.}{teal}
\litgroup{eval}{e1}{e3}{stageblue}{Model-level\\evaluation}
\litgroup{rag}{r1}{r5}{stagecyan}{Augmented\\LLM}
\litgroup{agent}{a1}{a5}{stagepurple}{LLM Agent}
\litgroup{mas}{d1}{d2}{teal}{Multi-Agent\\System}
\draw[treeline,line width=.45pt] (.105,0 |- e1.center)--(e1label.west);
\draw[treeline,line width=.45pt] (.105,0 |- e2.center)--(e2label.west);
\draw[treeline,line width=.45pt] (.105,0 |- e3.center)--(e3label.west);
\draw[treeline,line width=.45pt] (.105,0 |- r1.center)--(r1label.west);
\draw[treeline,line width=.45pt] (.105,0 |- r2.center)--(r2label.west);
\draw[treeline,line width=.45pt] (.105,0 |- r3.center)--(r3label.west);
\draw[treeline,line width=.45pt] (.105,0 |- r4.center)--(r4label.west);
\draw[treeline,line width=.45pt] (.105,0 |- r5.center)--(r5label.west);
\draw[treeline,line width=.45pt] (.105,0 |- a1.center)--(a1label.west);
\draw[treeline,line width=.45pt] (.105,0 |- a2.center)--(a2label.west);
\draw[treeline,line width=.45pt] (.105,0 |- a3.center)--(a3label.west);
\draw[treeline,line width=.45pt] (.105,0 |- a4.center)--(a4label.west);
\draw[treeline,line width=.45pt] (.105,0 |- a5.center)--(a5label.west);
\draw[treeline,line width=.45pt] (.105,0 |- d1.center)--(d1label.west);
\draw[treeline,line width=.45pt] (.105,0 |- d2.center)--(d2label.west);
\end{tikzpicture}
\caption{Literature map of model unlearning, evaluation, and system extensions. Categories overlap. $^*$: Supporting or boundary work; $^\dagger$: Preprint.}
\label{fig:literature-map}
\end{figure*}
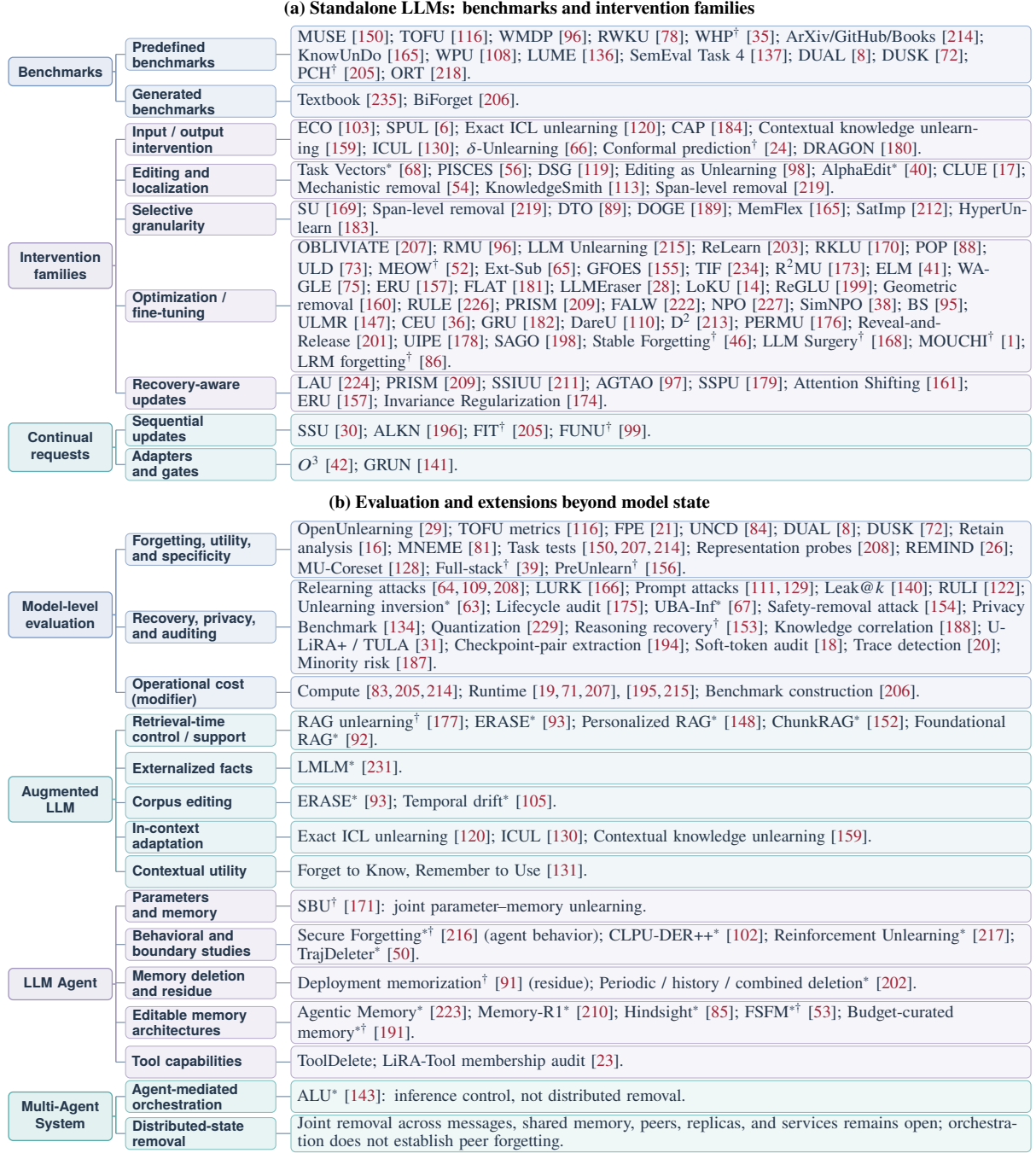

\endgroup

%% file: sections/11_open_challenges.tex
\section{Open Challenges and Future Directions}
\label{sec:challenges}

The evidence in Section~\ref{sec:trends} leaves five open problems, summarized in Table~\ref{tab:research-agenda}. For each, we examine why existing results fall short, what makes the problem difficult, and how progress could be tested. The objective is to determine when results from individual components support removal across the declared boundary and required paths.

\begin{table*}[!ht]
\caption{Open problems, sources of difficulty, and tests of progress.}
\label{tab:research-agenda}
\centering
\scriptsize
\setlength{\tabcolsep}{4pt}
\renewcommand{\arraystretch}{1.07}
\arrayrulecolor{stageblue!65}
\begin{tabularx}{\textwidth}{L{0.20\textwidth}L{0.36\textwidth}Y}
\toprule
\rowcolor{stageblue!16}
\textbf{Open problem} & \textbf{Why it remains difficult} & \textbf{Tests of progress} \\
\midrule
\cellcolor{stageblue!8}Defining removal targets & Shared knowledge; conflicting permissions & Adjudicated targets; preservation checks \\
\rowcolor{stageblue!4}
\cellcolor{stageblue!8}Tracing target influence & Missing lineage; transformed information & Known dependencies; localization errors \\
\cellcolor{stageblue!8}Coordinating interventions & Coupled state; delayed or inaccessible copies & Injected failures; end-to-end recovery \\
\rowcolor{stageblue!4}
\cellcolor{stageblue!8}Supporting broader claims & Finite probes; mismatched local guarantees & Adaptive attacks; explicit assumptions \\
\cellcolor{stageblue!8}Maintaining removal & Recurrence; cumulative damage and cost & Request streams; retests and total costs \\
\bottomrule
\end{tabularx}
\end{table*}

\smallskip \noindent \textbf{Defining Targets Without Removing Too Much.} Benchmark targets specify records, entities, or capabilities, but deployment requests can leave related knowledge and future use ambiguous. Removing one person's address, for example, does not necessarily authorize removing all knowledge of the same location. Editing, suppression, revocation, and deletion also produce different outcomes~\cite{nips/CooperEtAl25}. Shared knowledge and conflicting permissions make it difficult to specify which changes are required, which are permitted, and which would cause collateral damage.

Requests with adjudicated target and retain sets could clarify shared facts, derived summaries, and permitted historical uses. For memory or trajectory removal, the reference must specify which later decisions would differ without the target event; replay cannot simply assume that every other action remains unchanged. Tests should distinguish missed removal from unnecessary collateral change under these explicit choices. Unresolved authority conflicts must be reported alongside the target specification, rather than hidden by excluding disputed or inaccessible state.

\smallskip \noindent \textbf{Tracing Influence Across Different Forms of State.} Parametric localization methods identify useful neurons, directions, or logical neighborhoods~\cite{icml/GuoSSED25,emnlp/WangZYRRC25,acl/JeungYHKSHYN26}, while external stores can expose record-level dependencies. Neither fully traces influence across these different states. A summary may paraphrase a target without retaining its wording, and a learned skill may reflect an experience without storing a readable record. Consequently, literal matching misses some dependencies, whereas broad semantic matching can select unrelated material and increase collateral damage.

Testbeds with known source-to-summary and message-to-replica dependencies could measure both missed descendants and unnecessary selections. Controlled interventions could then test whether the identified state actually affects target disclosure or behavior. Learned tool-capability removal should be assessed separately from deletion of earlier tool outputs~\cite{icml/ChengA25}; memory utility after deletion likewise does not establish removal of behavioral influence~\cite{acl/XiongLXHLTLX26}. Provenance methods must balance tracing accuracy against storage and privacy costs: histories can preserve targets.

\smallskip \noindent \textbf{Coordinating Removal Across Components.} Retrieval control, knowledge-base editing, memory deletion, and tool unlearning demonstrate changes in particular components~\cite{corr/Wang2410RAG,naacl/LiLRZNA25,corr/Lei2606Agents,icml/ChengA25}. Combining them presents another problem: deleting a source before updating its cache can leave a disclosure window, while a disconnected replica can later restore the target. An acknowledgment may indicate receipt rather than verified execution. Local success therefore does not establish that the combined system meets the request, especially across independently controlled components.

Coordinated protocols should be tested with delayed messages, failed invalidations, and replica reconnection. Versioned records can establish which changes were applied, while end-to-end probes test whether surviving paths still reveal or use the target. Useful outcomes include residual influence during propagation, time until required checks pass, and total synchronization and retesting cost. When a required component cannot be inspected or changed, the result should identify the unresolved dependency instead of claiming complete removal from the requested system.

\smallskip \noindent \textbf{Connecting Finite Tests to Broader Claims.} Recovery through adversarial prompts, correlated facts, and checkpoint comparisons shows why one successful evaluation cannot establish universal removal~\cite{emnlp/ToL25,icml/ReisizadehRCPLH26,nips/Wei2025DoLR,nips/WuPLW25}. Expanding the probe set helps, but does not resolve unknown paths or attacks that adapt to the intervention. Combining local guarantees is also difficult when components use different references, access assumptions, and observation times. A model certificate cannot automatically cover retrieval or agent interactions beyond its assumptions.

Guarantees need explicit domains and checks of their deployment assumptions. Candidate approaches include bounds over declared access distributions, component contracts with explicit compatibility conditions, and audits linking inspected state to behavioral tests. Shared evaluation protocols~\cite{nips/DornaMZMKLM25} can support comparisons, provided they report uncertainty and distinguish a missing test from a failed one. Additional tests must justify broader coverage.

\smallskip \noindent \textbf{Maintaining Removal Under Repeated Updates.} Continual methods address successive requests through local updates, adapters, and selective activation~\cite{naacl/DouLLDW25,icml/wuerkaixi2025adaptive,corr/xu2026pch,iclr/gao25}. Yet preserving average utility across requests does not show that earlier targets remain removed. Later training, index rebuilding, and module consolidation can restore influence, while avoiding consolidation can accumulate storage and inference costs. The unresolved trade-off is therefore between durable removal, retained performance, and the resources needed to sustain both over a request stream.

Longitudinal benchmarks should retest earlier targets after subsequent requests and maintenance operations, while holding preservation criteria fixed. They should distinguish authorized reintroduction under $\tau_r$ from forbidden recurrence and report cumulative damage, storage, verification cost, and any retraining. Conflicting retention duties and uneven risk also require explicit approval and appeal procedures~\cite{icml/Wei25,nips/CooperEtAl25}. These procedures authorize changes; technical tests must establish their effectiveness and retained performance at the declared observation times.

Two milestones connect these directions: testing removal across known model, memory, and replica dependencies, then repeating those tests during request streams with injected failures. Together they assess whether local changes support a system claim and whether it survives later updates at an acceptable cost. Comparisons require fixed preservation criteria, declared observation horizons, and explicit disclosure of state beyond inspection or control.

%% file: sections/12_conclusion.tex
\section{Conclusion}
\label{sec:conclusion}

This survey uses a five-layer framework to compare removal requests, interventions, and the evidence supporting their outcomes. Model studies show that target construction, retained data, recovery tests, and score aggregation can change method comparisons; stronger suppression alone need not imply better removal. Agent studies extend the evidence to learned tool use, coordinated parameter and memory changes, and recoverable memory derivatives. These results support specific configurations, while joint removal across independent agents and later system updates remains unresolved. The seven-stage lifecycle and six evidence dimensions connect these findings to the additional tests broader claims require. Progress therefore depends on tracing affected state, testing propagation and restoration, and reporting retained performance alongside unresolved paths and observation horizons. Claims should follow the demonstrated results, with costs, control limits, and formal guarantees stated separately for each setting.